\documentclass[aps,prd,twocolumn,superscriptaddress,showpacs,floatfix]{revtex4-1}
\usepackage{graphicx,amsmath,amssymb,bm}
\begin{document}

\title{Nuclear structure aspects of spin-independent WIMP scattering off xenon}

\author{L.\ Vietze}
\email[E-mail:~]{lvietze@theorie.ikp.physik.tu-darmstadt.de}
\affiliation{Institut f\"ur Kernphysik, 
Technische Universit\"at Darmstadt, 
64289 Darmstadt, Germany}
\affiliation{ExtreMe Matter Institute EMMI, 
GSI Helmholtzzentrum f\"ur Schwerionenforschung GmbH, 
64291 Darmstadt, Germany}
\author{P.\ Klos}
\email[E-mail:~]{pklos@theorie.ikp.physik.tu-darmstadt.de}
\affiliation{Institut f\"ur Kernphysik, 
Technische Universit\"at Darmstadt, 
64289 Darmstadt, Germany}
\affiliation{ExtreMe Matter Institute EMMI, 
GSI Helmholtzzentrum f\"ur Schwerionenforschung GmbH, 
64291 Darmstadt, Germany}
\author{J.\ Men\'{e}ndez}
\email[E-mail:~]{menendez@nt.phys.s.u-tokyo.ac.jp}
\altaffiliation{\\Present address: Department of Physics, University of Tokyo, Hongo, Tokyo 113-0033, Japan}
\affiliation{Institut f\"ur Kernphysik, 
Technische Universit\"at Darmstadt, 
64289 Darmstadt, Germany}
\affiliation{ExtreMe Matter Institute EMMI, 
GSI Helmholtzzentrum f\"ur Schwerionenforschung GmbH, 
64291 Darmstadt, Germany}
\author{W.\ C.\ Haxton}
\email[E-mail:~]{haxton@berkeley.edu}
\affiliation{Department of Physics, University of California, 
Berkeley, CA 94720, USA}
\affiliation{Lawrence Berkeley National Laboratory, Berkeley, CA 94720, USA}
\author{A.\ Schwenk}
\email[E-mail:~]{schwenk@physik.tu-darmstadt.de}
\affiliation{Institut f\"ur Kernphysik, 
Technische Universit\"at Darmstadt, 
64289 Darmstadt, Germany}
\affiliation{ExtreMe Matter Institute EMMI, 
GSI Helmholtzzentrum f\"ur Schwerionenforschung GmbH, 
64291 Darmstadt, Germany}

\begin{abstract}
We study the structure factors for spin-independent WIMP scattering
off xenon based on state-of-the-art large-scale shell-model
calculations, which are shown to yield a good spectroscopic
description of all experimentally relevant isotopes. Our results are
based on the leading scalar one-body currents only. At this level and
for the momentum transfers relevant to direct dark matter detection,
the structure factors are in very good agreement with the
phenomenological Helm form factors used to give experimental limits
for WIMP-nucleon cross sections. In contrast to spin-dependent WIMP
scattering, the spin-independent channel, at the one-body level, is
less sensitive to nuclear structure details. In addition, we
explicitly show that the structure factors for inelastic scattering
are suppressed by $\sim 10^{-4}$ compared to the coherent elastic
scattering response. This implies that the detection of inelastic
scattering will be able to discriminate clearly between
spin-independent and spin-dependent scattering. Finally, we provide
fits for all calculated structure factors.
\end{abstract}


\maketitle

\section{Introduction}

About $27\%$ of the energy density in the universe consists of dark
matter that rarely interacts with baryonic matter~\cite{Feng}. Weakly
interacting massive particles (WIMPs), postulated by supersymmetric
extensions of the standard model, are among the most promising dark
matter candidates, as their predicted density would naturally account
for the observed dark matter density~\cite{Bertone}. Furthermore,
WIMPs interact with quarks, allowing for direct dark matter detection
by the observation of the nuclear recoil induced by WIMP scattering
off nuclei~\cite{Baudis}. Several experiments worldwide are searching
for this dark matter signature~\cite{CDMSII,EdelweissII,Simple,ZEPLINIII,%
XenonSI,LUX}, but so far no unambiguous detection has been achieved.
In addition, WIMPs could also scatter inelastically~\cite{Ellis},
thereby exciting the nucleus and yielding a different dark matter signal.

The analysis of direct detection experiments requires knowledge of the
nuclear structure factors. For a given coupling between WIMPs and
nucleons, these encode the nuclear structure aspects relevant for
WIMP-nucleus scattering.  In this work, we calculate the structure
factors for spin-independent (SI) WIMP scattering, complementing our
previous work on elastic~\cite{Menendez,Klos} and
inelastic~\cite{inelastic} spin-dependent (SD) interactions.  We focus
on scattering off xenon, which is used as target of major direct
detection experimental efforts such as XENON and
LUX~\cite{XenonSI,LUX}.

For the SI WIMP coupling to nucleons, we take the standard
scalar-isoscalar current-current interaction Lagrangian as discussed
in Ref.~\cite{Engel}.  In addition, a reliable description of the
nuclear states involved in the scattering process is needed.  In this
work, we perform state-of-the-art nuclear structure calculations,
which take advantage of progress in nuclear interactions and computing
capabilities, and compare our results to the phenomenological
structure factors typically used in dark matter detection analyses and
to other calculations.

This paper is organized as follows.  In Sec.~\ref{Sec:SI_interaction}
we discuss the structure factor for SI scattering starting from the
effective WIMP-nucleon interaction Lagrangian, considering the leading
scalar one-body currents.  Our nuclear structure calculations are
discussed in Sec.~\ref{Sec:spectra}. In Sec.~\ref{sec:structure factors}
we present the resulting structure factors for elastic SI WIMP
scattering off all stable xenon isotopes. These are compared in
Sec.~\ref{Sec:Comparison Helm} to the phenomenological Helm form
factors used in most analyses of direct detection experiments.  We
also compare our results to the recent calculations of Fitzpatrick
{\it et al.}~\cite{Fitzpatrick}, for both SI and SD cases.  Inelastic
WIMP scattering off xenon and its capability to distinguish between SI
and SD interactions are discussed in Sec.~\ref{Sec:Spin-independent}.
Finally, we summarize in Sec.~\ref{Sec:Summary}.

\section{Spin-independent WIMP-nucleus scattering}	
\label{Sec:SI_interaction}

The SI interaction of WIMPs with nuclei, assuming spin 1/2 neutralinos,
is described by the low-momentum-transfer Lagrangian~\cite{Engel}
\begin{equation}
\label{eq:Lagrangian}
\mathcal{L}_\chi^\text{SI} = \frac{G_F}{\sqrt{2}} \int d^3{\bf r} \, 
j({\bf r}) \, S({\bf r}) \,,
\end{equation}
where $G_F$ is the Fermi coupling constant, and $j({\bf r})$ and
$S({\bf r})$ denote the scalar leptonic and the scalar hadronic
current, respectively.  The leptonic current is given by kinematics of
the WIMP field $\chi$,
\begin{equation}
j({\bf r}) = \overline{\chi} \chi = 
\delta_{s_f,s_i} \, e^{-i \mathbf{{\bf q}\cdot{\bf r}}} \,,
\end{equation}
where $s_f$, $s_i =\pm 1/2$ are the final and initial spin projections
of the WIMP and $\mathbf{q}$ is the momentum transfer from nucleons to
neutralinos.  As in Ref.~\cite{Engel}, we take the hadronic current of
the nucleons to be purely isoscalar with coupling $c_0$. We take into
account only the leading one-body currents, so that the scalar
nuclear current is a sum over single nucleons,
\begin{equation}
S({\bf r})=c_0\sum_{i=1}^A \delta^{(3)}({\bf r}-{\bf r}_i) \,.
\end{equation}
However, additional contributions enter from two-body
currents~\cite{Prezeau,Cirigliano}, whose importance is under
discussion~\cite{Beane}.

The differential cross section for SI WIMP scattering off a nucleus
with initial state $|i\rangle$ and final state $|f\rangle$ is obtained
from the Lagrangian density of Eq.~(\ref{eq:Lagrangian})~\cite{Engel}:
\begin{align}
\frac{d\sigma}{dq^2} &= \frac{2}{(2J_i+1)\pi v^2}
\sum_{s_f,s_i}\sum_{M_f,M_i} \bigl|\langle f|\mathcal{L}_\chi^\text{SI}|i\rangle
\bigr|^2 \nonumber\\
&=\frac{8 G_F^2}{(2J_i+1)v^2} \, S_S(q) \,.
\end{align}	
The total angular momentum of the initial and final states of the
nucleus are denoted by $J_i$ and $J_f$, with projections $M_i$ and
$M_f$, $v$ is the WIMP velocity, and $S_S(q)$ is the scalar structure
factor.  As the target is unpolarized, one averages over initial
projections and sums over the final ones.  Following
Ref.~\cite{Walecka}, the structure factor can be decomposed as a sum
over multipoles ($L$) of the reduced matrix elements of the Coulomb
projection $\mathcal{C}_L$ of the scalar current:
\begin{equation}
S_S(q) = \sum_{L=0}^{\infty} \bigl| \langle J_f \lVert \mathcal{C}_L(q) 
\rVert \, J_i \rangle \bigr|^2 \,,
\end{equation}
with
\begin{equation}
\mathcal{C}_{LM}(q) = c_0 \sum_{i=1}^{A} j_L (qr_i) Y_{LM} ({\bf r}_i) \,.
\label{Eq:multipole}
\end{equation}

Each Coulomb multipole in Eq.~(\ref{Eq:multipole}) has a given parity set by
the spherical harmonic, $\Pi(Y_{LM})=(-1)^L$.  For elastic scattering
the initial and final states are the same and
$J_i^{\Pi_i}=J_f^{\Pi_f}$, so that only even $L$ multipoles
contribute. For inelastic scattering the parity of the initial and
final states can differ, and the allowed multipoles are given by
\begin{align}
\Pi_f &= \Pi_i \Rightarrow L \text{ even} \,, \\[1mm]
\Pi_f &\neq \Pi_i \Rightarrow L \text{ odd} \,.
\end{align}
Note that the odd $L$ multipoles in elastic scattering are also
forbidden by time-reversal symmetry.

\section{Spectra of even-mass xenon isotopes}
\label{Sec:spectra}

Xenon has proton number $Z=54$, and the neutron number of the stable
isotopes ranges from $N=74-82$.  Our calculations assume an isospin
symmetric $^{100}$Sn core.  For the remaining nucleons we consider a
valence space consisting of the 0g$_{7/2}$, 1d$_{5/2}$, 1d$_{3/2}$,
2s$_{1/2}$, and 0h$_{11/2}$ orbitals, both for neutrons and protons,
and the effective nuclear interaction
GCN5082~\cite{Caurier,MenendezNPA}. The same valence space and
nuclear interaction have been used for the study of SD WIMP scattering
off the odd-mass isotopes $^{129}$Xe and
$^{131}$Xe~\cite{Menendez,Klos,inelastic} and for the neutrinoless double-beta
decay of $^{136}$Xe~\cite{Caurier,MenendezNPA}. Throughout all
calculations we use the shell-model code ANTOINE~\cite{Antoine,ISM}.

The even-mass isotopes $^{132}$Xe, $^{134}$Xe, and $^{136}$Xe are
calculated by exact diagonalization in the valence space.  However,
for $^{128}$Xe and $^{130}$Xe, proton and neutron excitations from the
0g$_{7/2}$ and 1d$_{5/2}$ into the 1d$_{3/2}$, 2s$_{1/2}$, and
0h$_{11/2}$ orbitals were restricted to three and six, respectively,
to keep the matrix dimension tractable. These truncations should not
affect the most important shell-model configurations. The matrix
dimensions associated with the nuclear structure calculations of the
even-mass xenon isotopes are given in Table~\ref{tab:dimensions}.

\begin{table}[b]
\begin{center}
\caption{Matrix dimensions for the even-mass xenon isotopes. For 
$^{132}$Xe, $^{134}$Xe, and $^{136}$Xe the dimension is that of the
full valence space, whereas for $^{128}$Xe and $^{130}$Xe the
calculations are restricted (as discussed in the text).
\label{tab:dimensions}}
\begin{tabular*}{0.48\textwidth}{l|c|c|c|c|r}
\hline \hline
Isotope	& \multicolumn{1}{c|}{$^{128}$Xe} & \multicolumn{1}{c|}{$^{130}$Xe} & \multicolumn{1}{c|}{$^{132}$Xe} & \multicolumn{1}{c|}{$^{134}$Xe} & \multicolumn{1}{r}{$^{136}$Xe} \\
\hline
Dimension & \,$373 \cdot 10^{6}$ \,& \,$410 \cdot 10^{6}$\, & \,
$21 \cdot 10^{6}$ \,& \,$335 \cdot 10^{3}$\, &\, $1500$\\
\hline \hline
\end{tabular*}
\end{center}
\end{table}
	
\begin{figure}[!tb] 
\begin{center}
\includegraphics[width=\columnwidth,clip=]{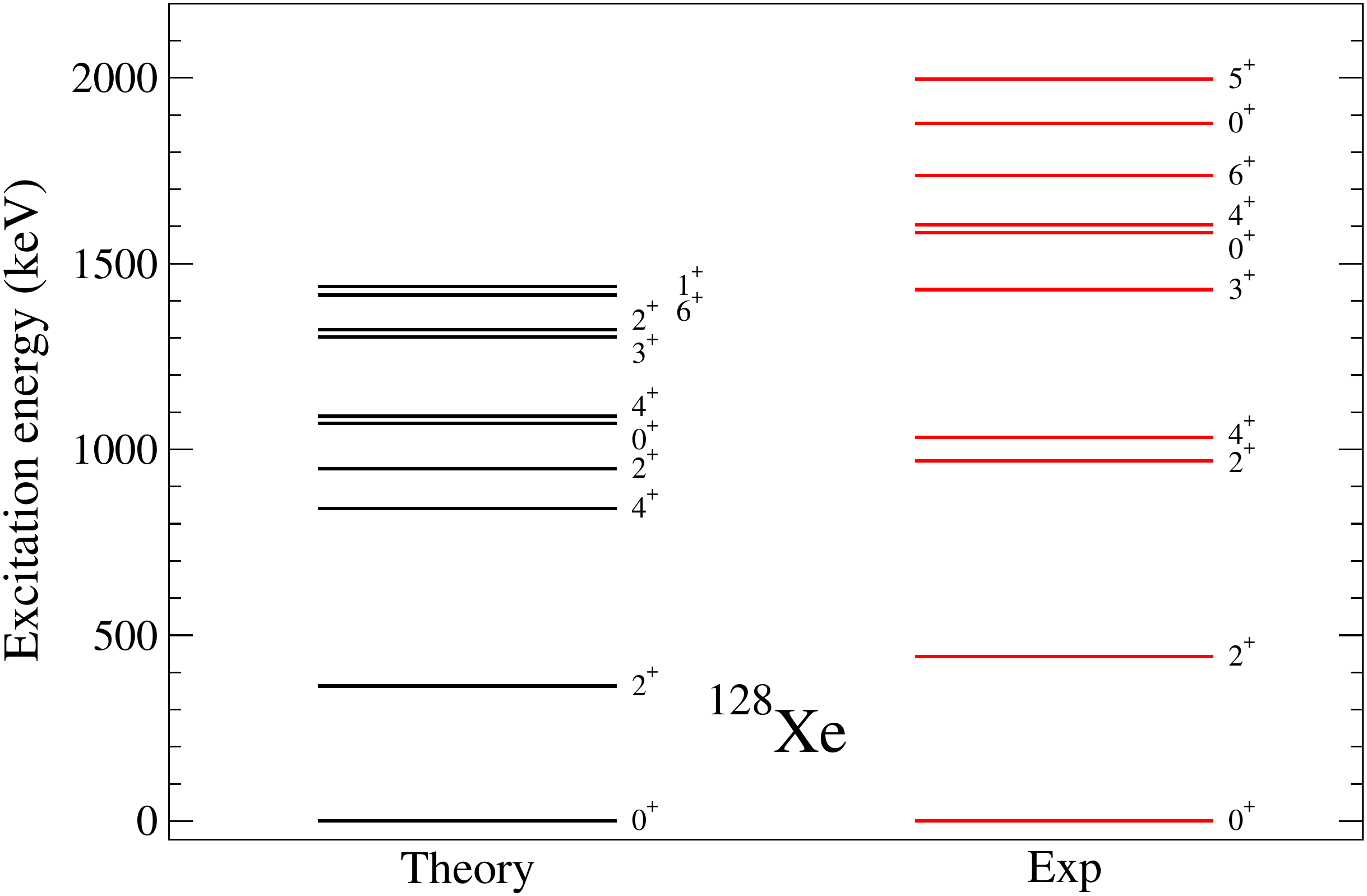}
\end{center}
\caption{Comparison of the calculated spectrum of $^{128}$Xe with 
experiment~\cite{nndc}. The calculation is performed in a restricted 
valence space (as discussed in the text).\label{fig:Xe128-spec}}
\end{figure}

\begin{figure}[!tb]
\begin{center}
\includegraphics[width=\columnwidth,clip=]{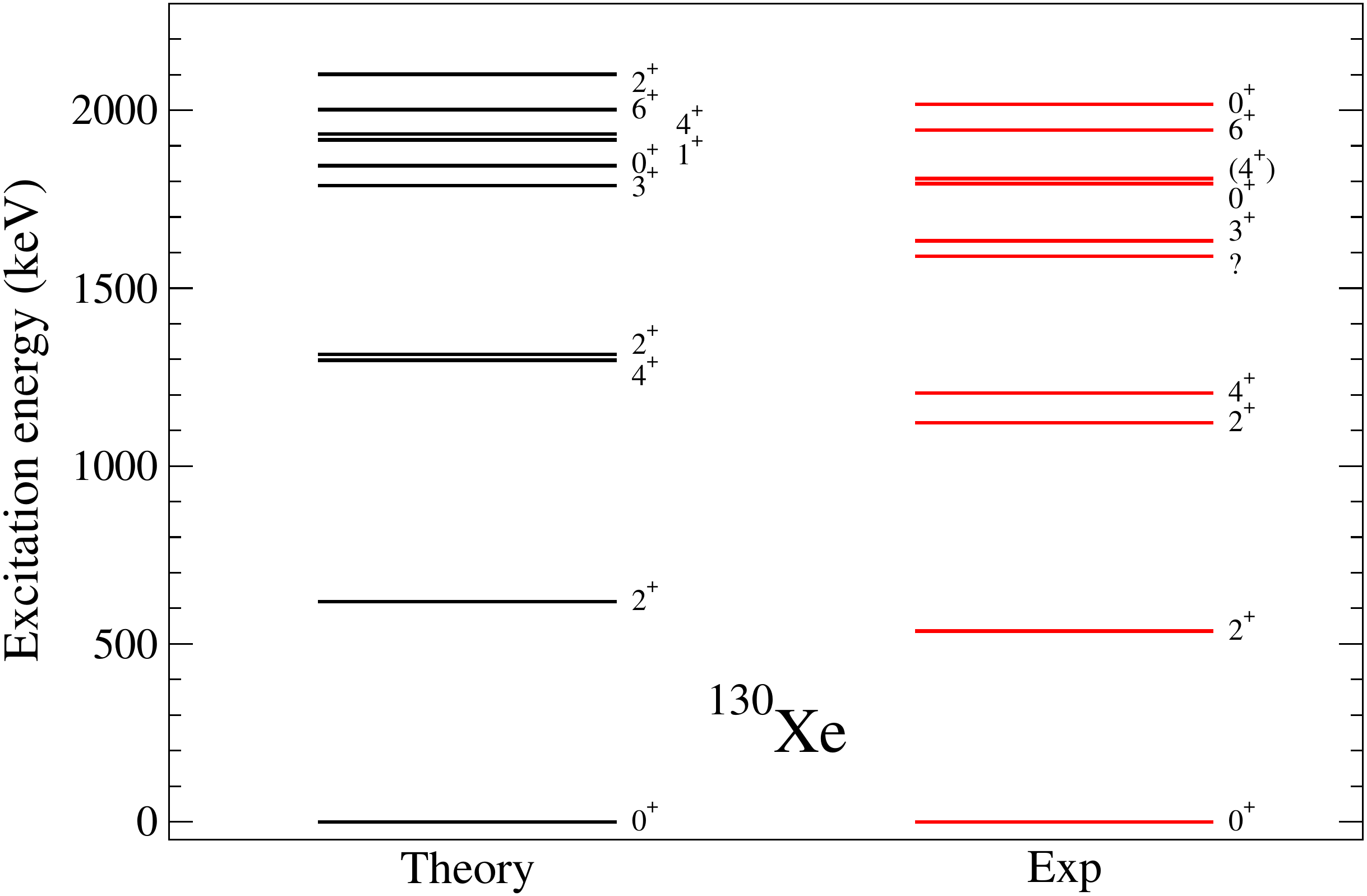}
\caption{Comparison of the calculated spectrum of $^{130}$Xe with 
experiment~\cite{nndc}. The calculation is performed in a restricted
valence space (as discussed in the text).\label{fig:Xe130-spec}}
\end{center}
\end{figure}

\begin{figure}[!htb]
\begin{center}
\includegraphics[width=\columnwidth,clip=]{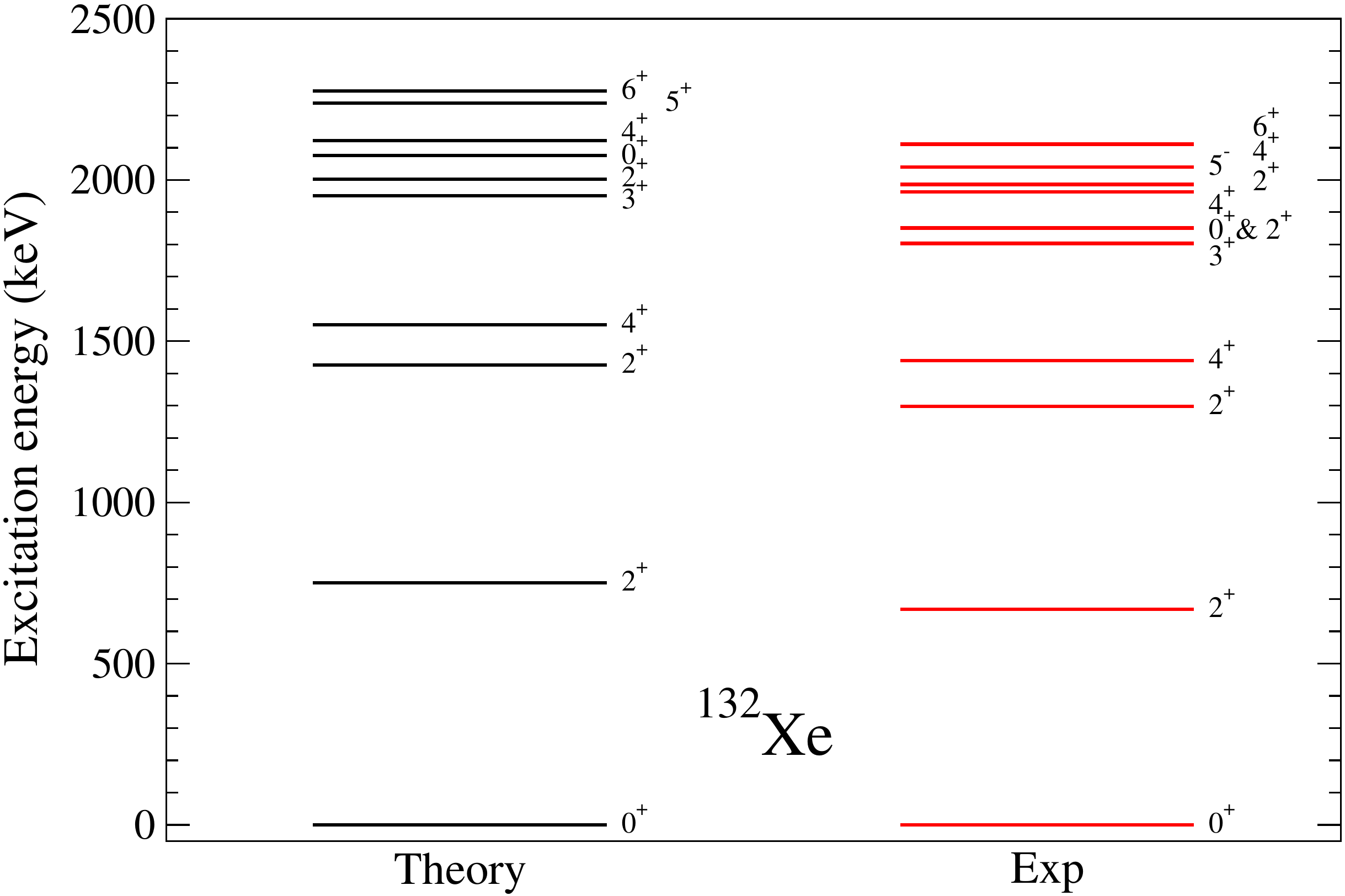}
\end{center}
\caption{Comparison of the calculated spectrum of $^{132}$Xe with 
experiment~\cite{nndc}. The calculation is performed in the full 
valence space.\label{fig:Xe132-spec}}
\end{figure}

\begin{figure}[!htb]
\begin{center}
\includegraphics[width=\columnwidth,clip=]{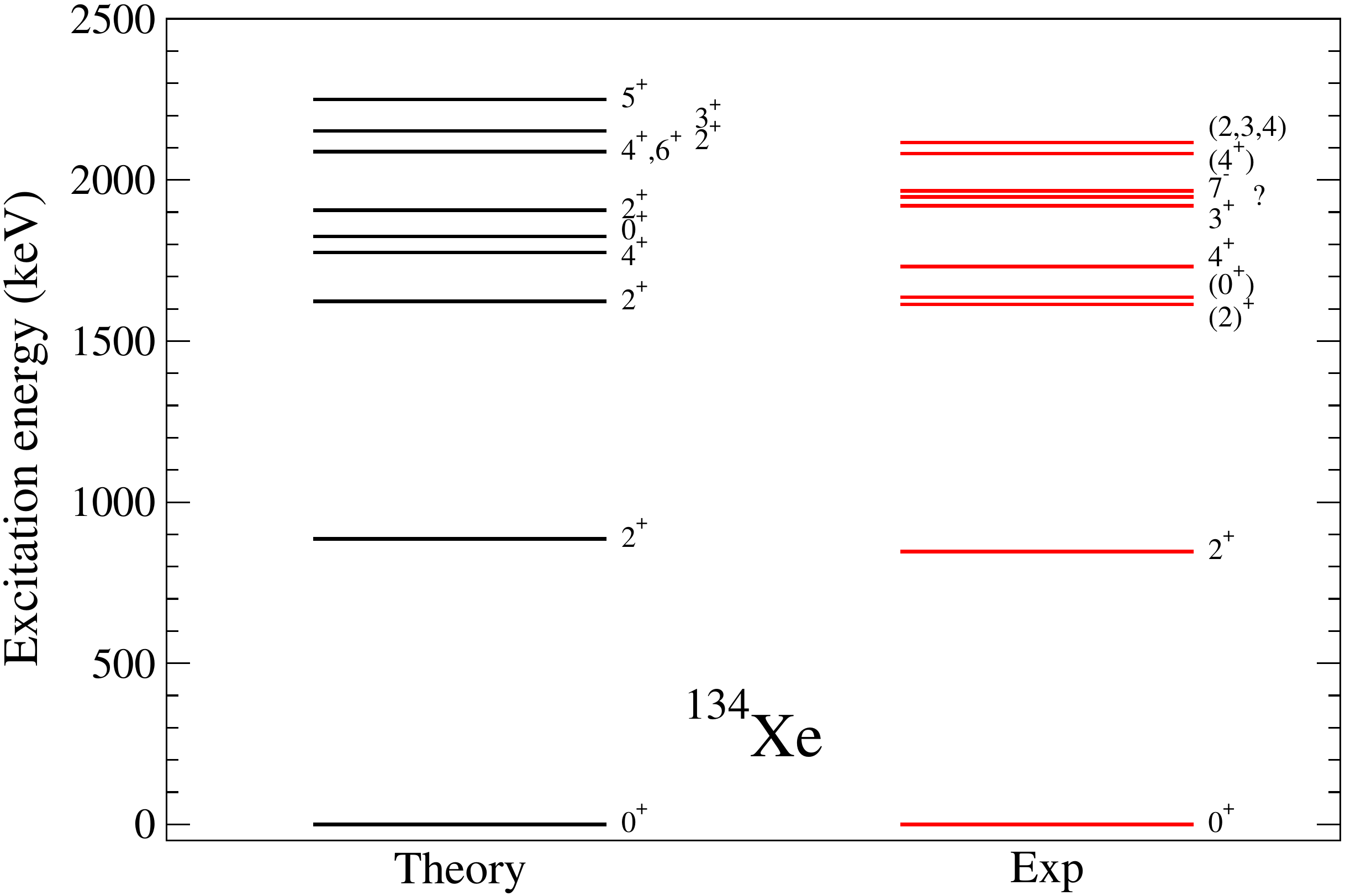}
\end{center}
\caption{Same as Fig.~\ref{fig:Xe132-spec} but for $^{134}$Xe.
\label{fig:Xe134-spec}}
\end{figure}

\begin{figure}[!htb]
\begin{center}
\includegraphics[width=\columnwidth,clip=]{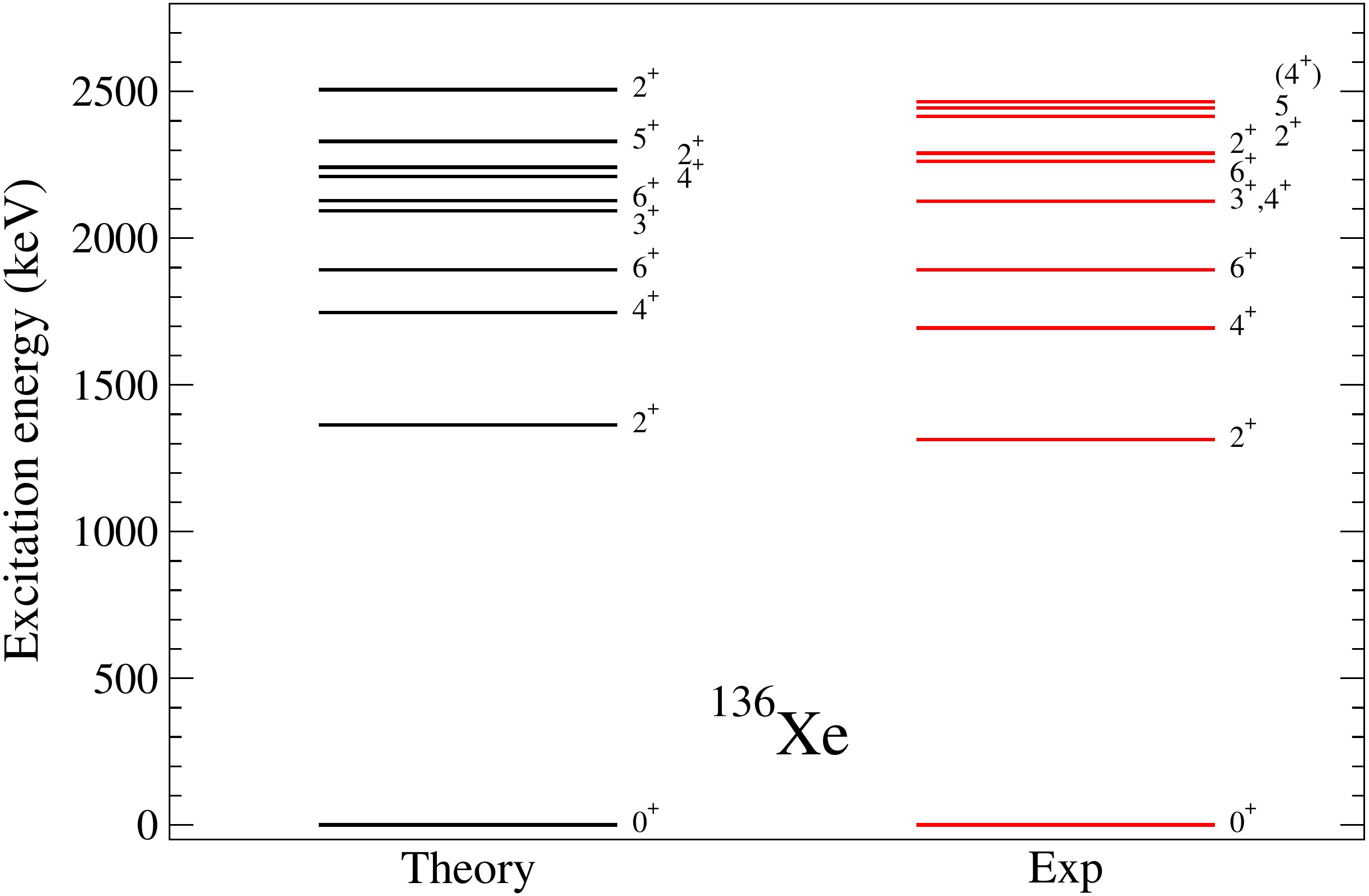}
\end{center}
\caption{Same as Fig.~\ref{fig:Xe132-spec} but for $^{136}$Xe.
\label{fig:Xe136-spec}}
\end{figure}

Figures~\ref{fig:Xe128-spec}--\ref{fig:Xe136-spec} compare the
calculated spectra of the stable even-mass xenon isotopes to
experiment. The ten lowest lying states are given. The spectra of
$^{129}$Xe and $^{131}$Xe were shown and discussed in
Ref.~\cite{Klos}. The overall agreement with experiment is very good
in all cases.  The spin/parity and location of the first excited
$2^+_1$ state is very well reproduced along the isotopic chain, and
the spacing between this state and the following excited states is
also in very good agreement with experiment.

In $^{128}$Xe, $^{130}$Xe, and $^{132}$Xe, the second and third
excited states form a doublet of $2^+_2$, $4^+_1$ states, well
separated from other excited states.  This situation is well reproduced
in our calculations of $^{130}$Xe and $^{132}$Xe, but not for
$^{128}$Xe, where the calculated spectra is significantly more
compressed than experiment. This disagreement is due to the
restrictions imposed on the valence space for the $^{128}$Xe
calculations. These truncations mainly affect higher lying states
above $1$~MeV, so that one can be confident in the calculation of the
structure factor for WIMP scattering, as this involves only the ground
state.

For $^{134}$Xe and $^{136}$Xe, the second and higher excited states
are relatively close to each other (especially in $^{134}$Xe), and the
location and spin/parity of those states obtained in our calculations
in the full valence space is in good agreement to the (sometimes
tentative) experimental assignments.

\section{Structure factors for spin-independent WIMP scattering off xenon
isotopes}
\label{sec:structure factors}

\begin{figure}[!htb]
\begin{center}
\includegraphics[width=\columnwidth,clip=]{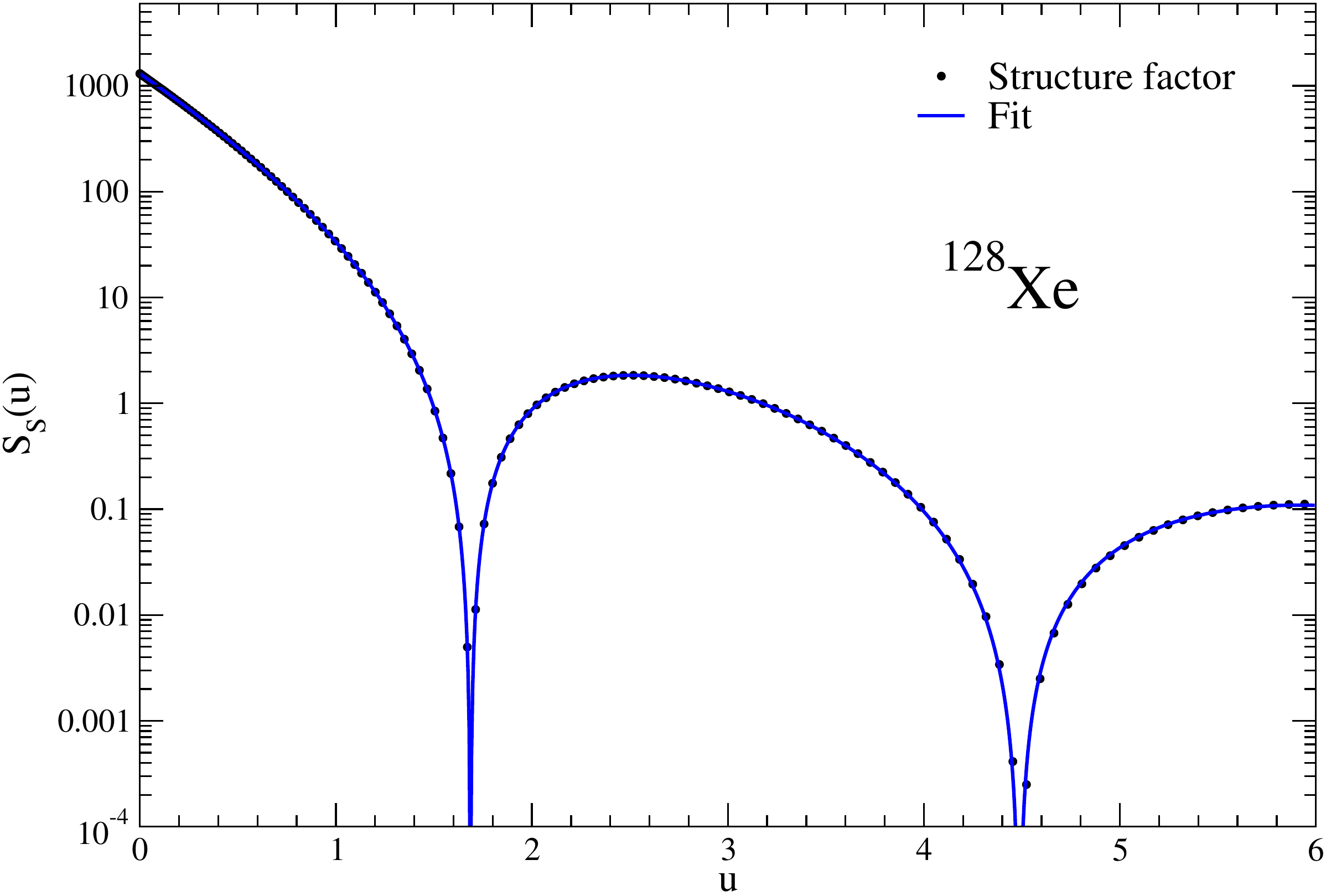}
\end{center}
\caption{(color online). Structure factor $S_S(u)$ for $^{128}$Xe 
(black dots) with a fit (solid blue line) given in Table~\ref{tab:fits}.
\label{fig:Xe128_sf_fit}}
\end{figure}

\begin{figure}[!htb]
\begin{center}
\includegraphics[width=\columnwidth,clip=]{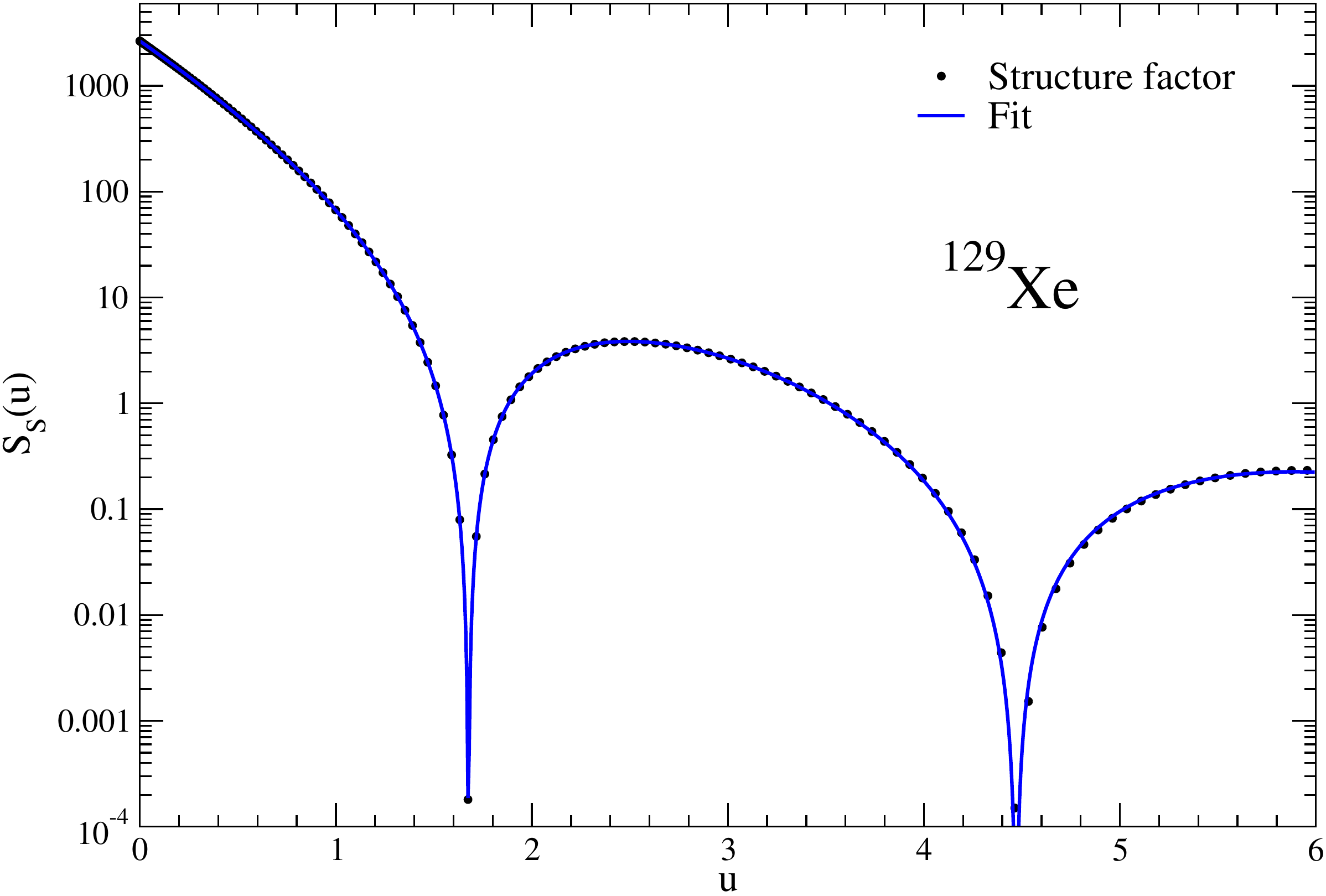}
\end{center}
\caption{(color online). Same as Fig.~\ref{fig:Xe128_sf_fit} but for
$^{129}$Xe.
\label{fig:Xe129_sf_fit}}
\end{figure}

\begin{figure}[!htb]
\begin{center}
\includegraphics[width=\columnwidth,clip=]{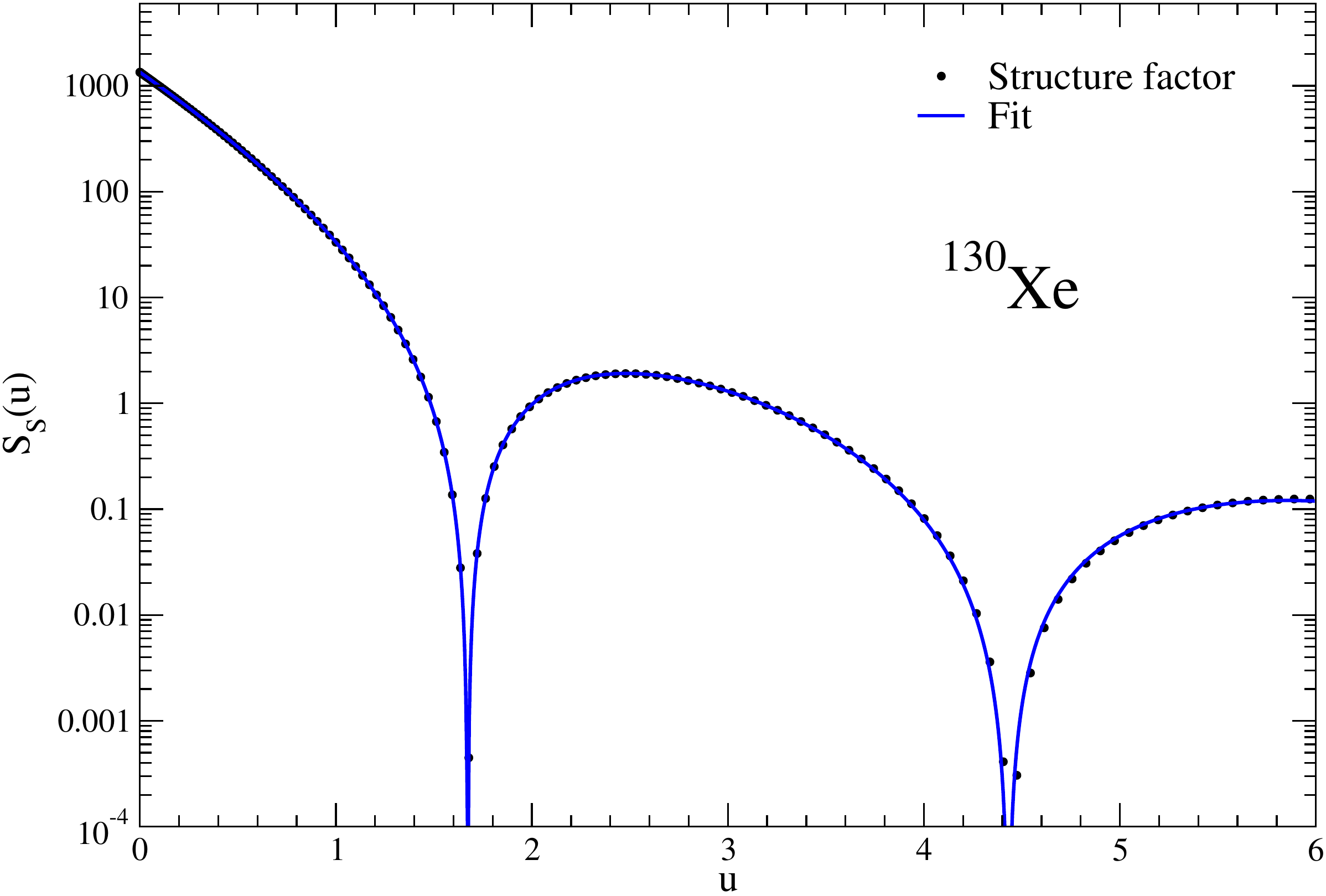}
\end{center}
\caption{(color online). Same as Fig.~\ref{fig:Xe128_sf_fit} but for
$^{130}$Xe.
\label{fig:Xe130_sf_fit}}
\end{figure}

\begin{figure}[!htb]
\begin{center}
\includegraphics[width=\columnwidth,clip=]{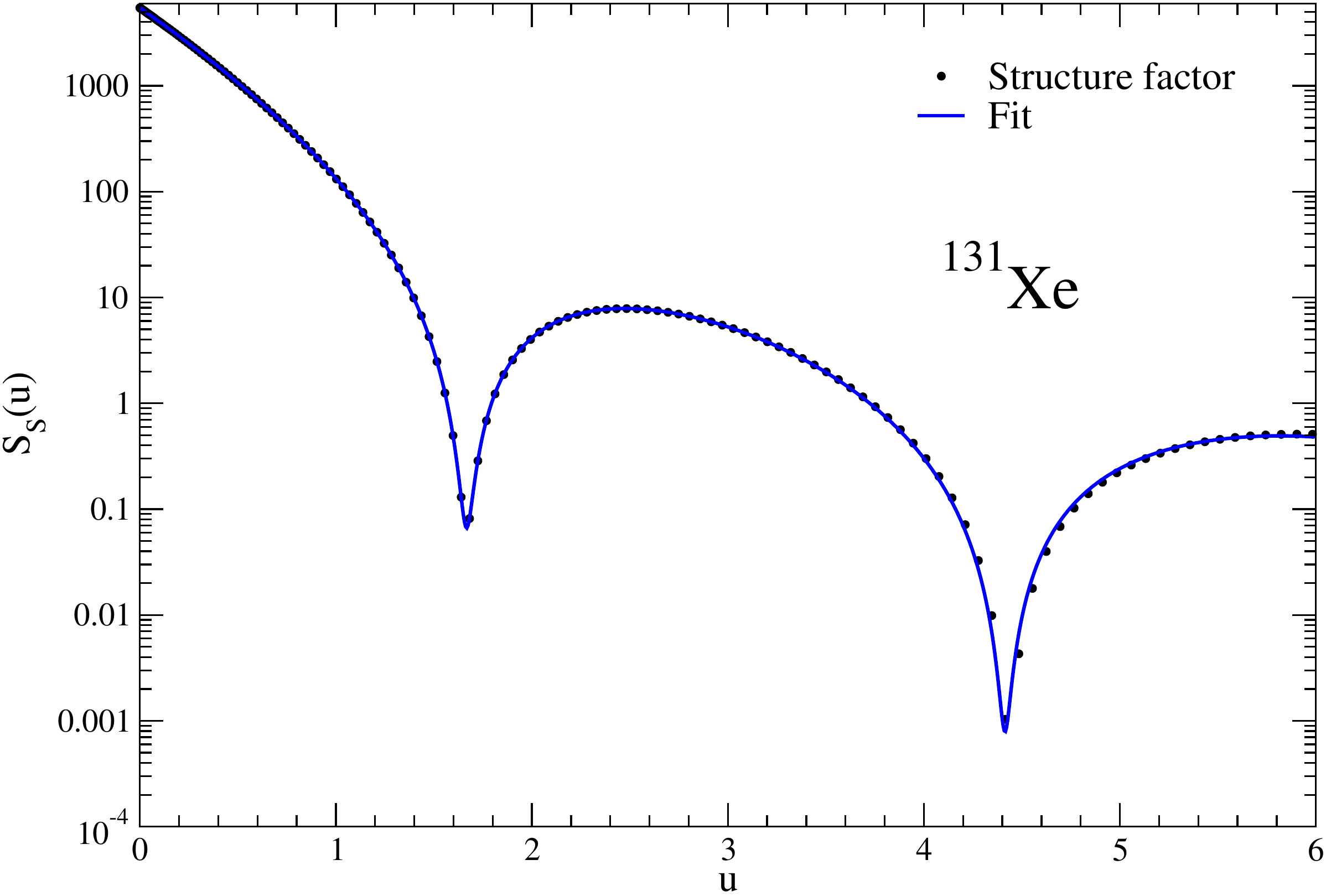}
\end{center}
\caption{(color online). Same as Fig.~\ref{fig:Xe128_sf_fit} but for
$^{131}$Xe.
\label{fig:Xe131_sf_fit}}
\end{figure}

\begin{figure}[!htb]
\begin{center}
\includegraphics[width=\columnwidth,clip=]{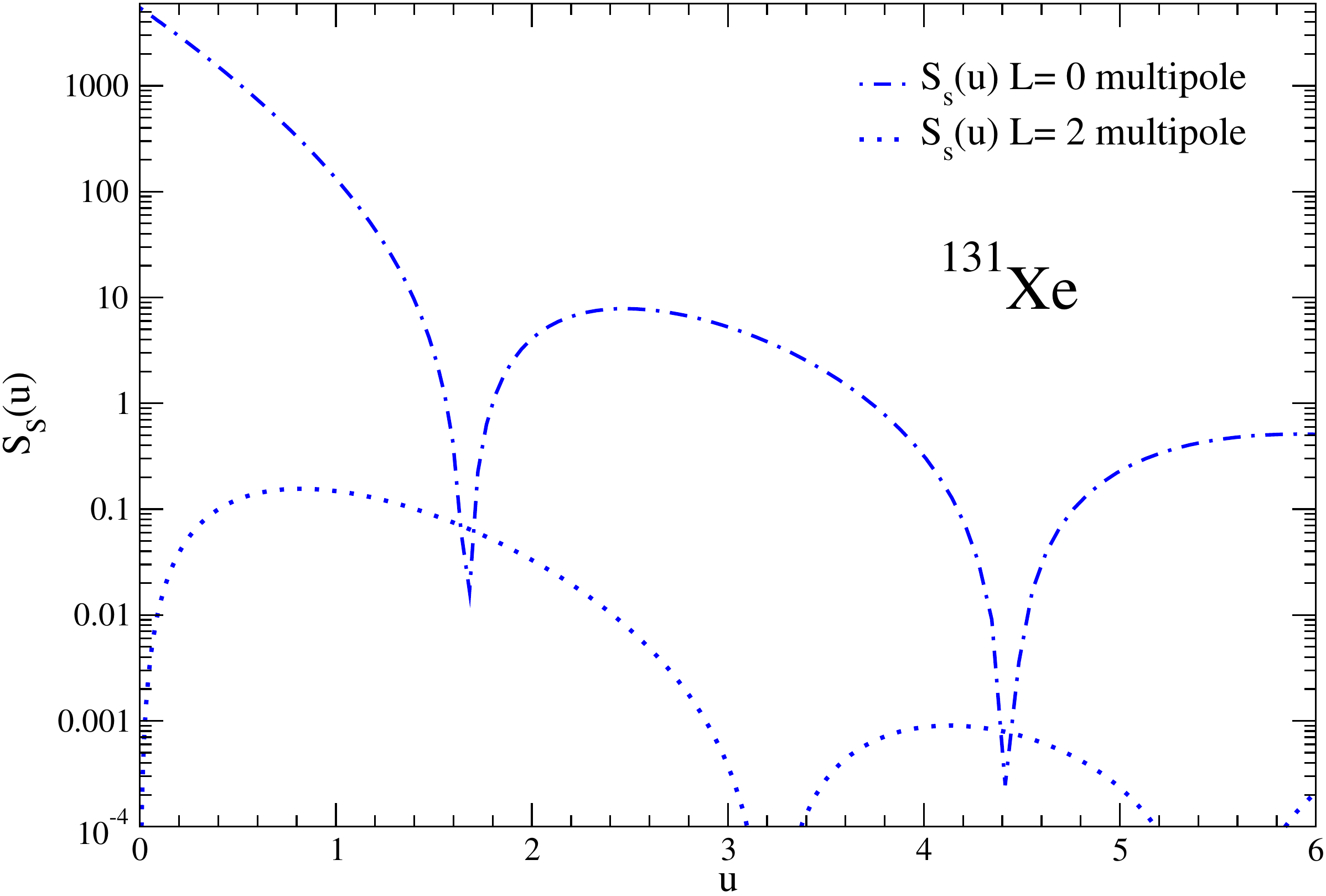}
\end{center}
\caption{(color online). Decomposition of the structure factor 
$S_S(u)$ for $^{131}$Xe in $L=0$ (dashed-dotted line) and $L=2$ 
(dotted line) multipoles.\label{fig:Xe131_sf_l}}
\end{figure}

\begin{figure}[!htb]
\begin{center}
\includegraphics[width=\columnwidth,clip=]{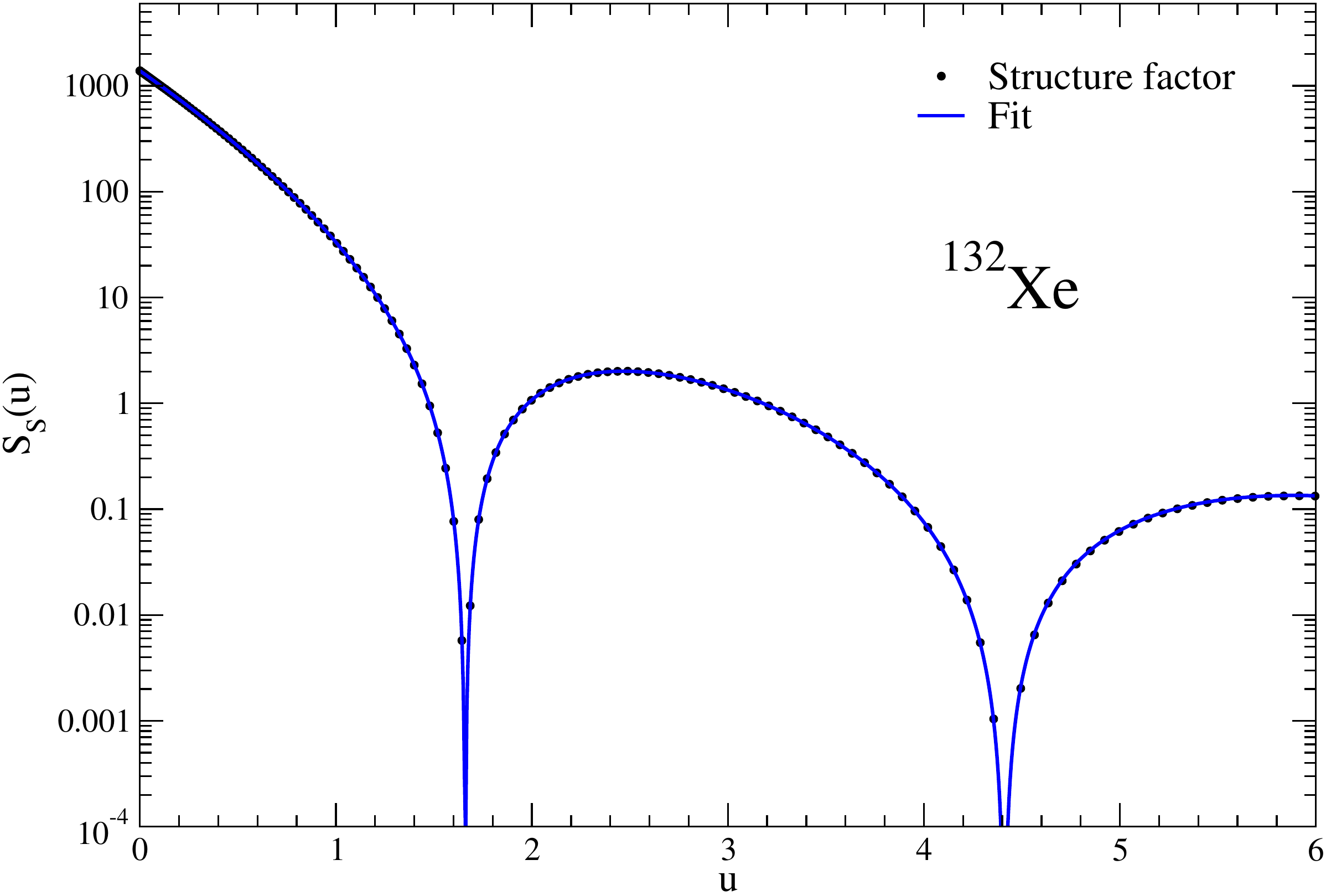}
\end{center}
\caption{(color online). Same as Fig.~\ref{fig:Xe128_sf_fit} but for
$^{132}$Xe.
\label{fig:Xe132_sf_fit}}
\end{figure}

\begin{figure}[htb]
\begin{center}
\includegraphics[width=\columnwidth,clip=]{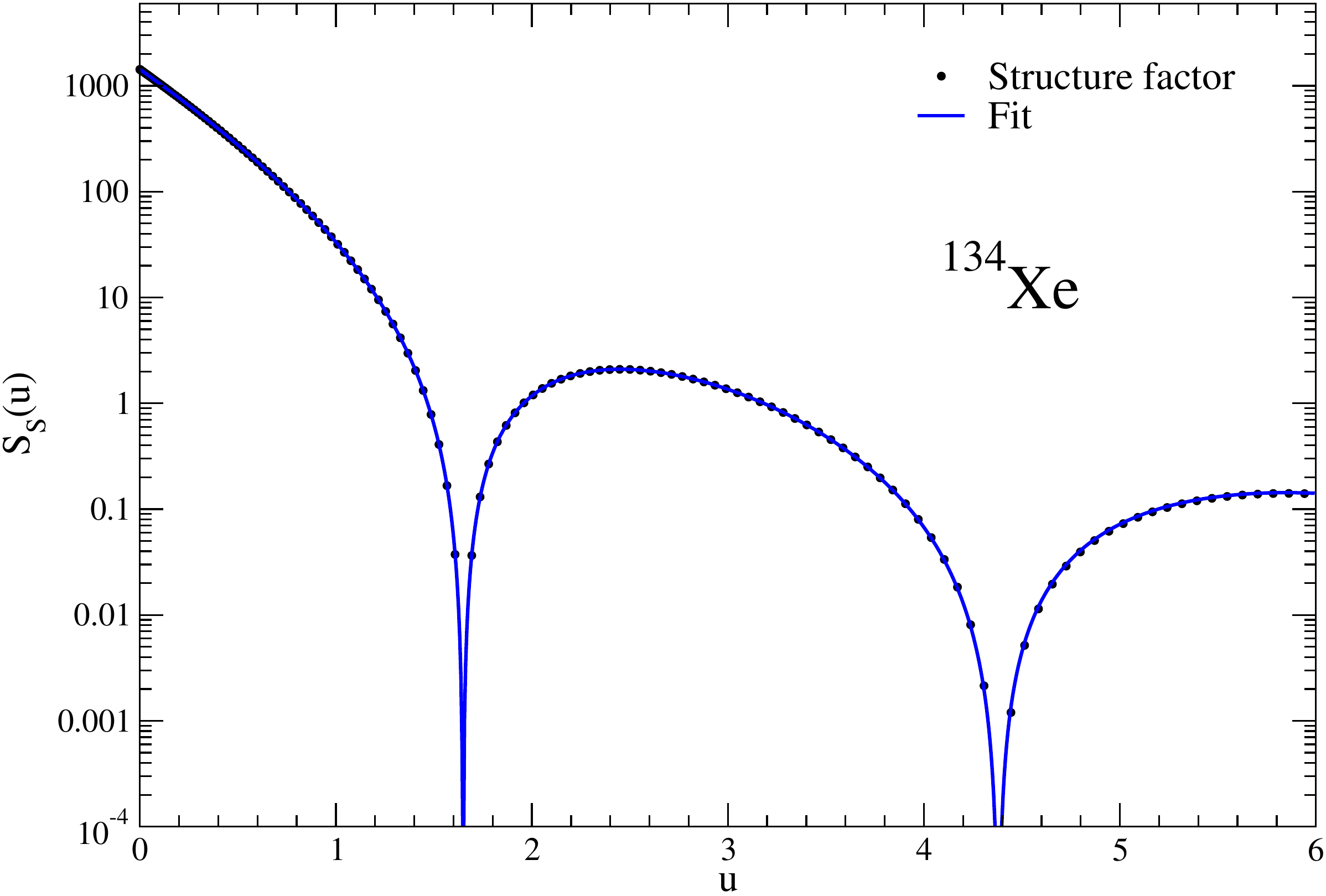}
\end{center}
\caption{(color online). Same as Fig.~\ref{fig:Xe128_sf_fit} but for
$^{134}$Xe.
\label{fig:Xe134_sf_fit}}
\end{figure}

\begin{figure}[htb]
\begin{center}
\includegraphics[width=\columnwidth,clip=]{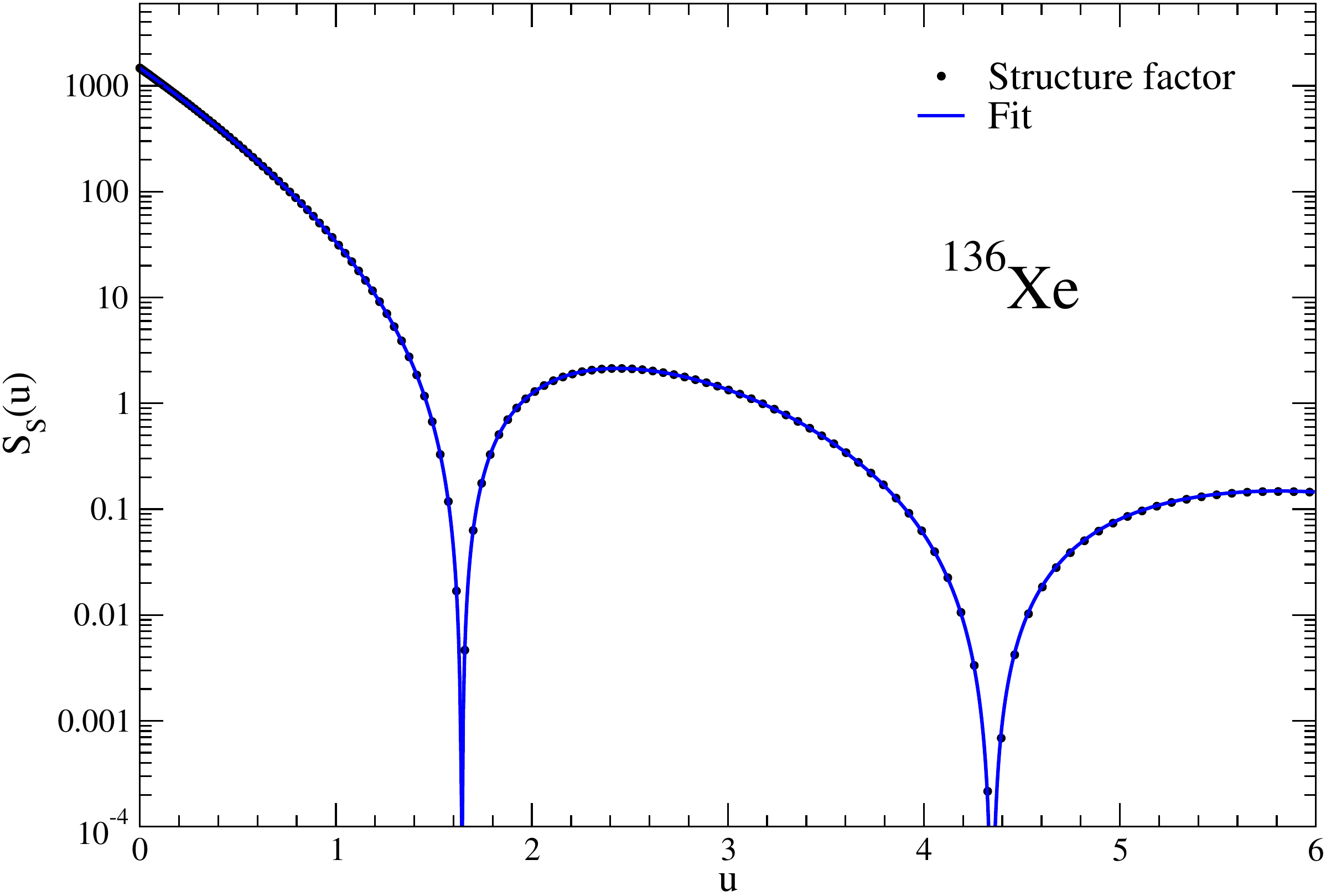}
\end{center}
\caption{(color online). Same as Fig.~\ref{fig:Xe128_sf_fit} but for
$^{136}$Xe.
\label{fig:Xe136_sf_fit}}
\end{figure}

The resulting structure factors for elastic ($J_i=J_f=J$) SI WIMP
scattering off the seven stable xenon isotopes are shown in
Figs.~\ref{fig:Xe128_sf_fit}--\ref{fig:Xe136_sf_fit}.  The structure
factors $S_S(q)$ are plotted as a function of the dimensionless
variable $u = q^2 b^2/2$, where $q$ is the momentum transfer and $b$
is the harmonic-oscillator length, defined as $b=\sqrt{\hbar/m
\omega}$ with $m$ the nucleon mass and $\omega$ the oscillator frequency, 
taken as $\hbar \omega = (45 A^{-1/2}-25 A^{-2/3})$~MeV.

At zero momentum transfer, $S_S(0)$ receives contributions only from
the $L=0$ multipole and is model-independent:
\begin{equation}
S_S(0)= A^2 \, c_0^2 \, \frac{2J+1}{4 \pi} \,.
\end{equation}
This reflects the well-known coherence of the contributions of all $A$
nucleons in SI scattering. Consequently, near $u=0$ the spin-averaged
structure factors are essentially identical for all xenon isotopes,
apart from small variations in $A^2$.

\begin{table*}[t]
\caption{Spin/parity $J^{\Pi}$, harmonic-oscillator length $b$, and fit 
coefficients for the structure factors $S_S(u)$ corresponding to
$S_S(u) = \frac{2J+1}{4 \pi} \, e^{-u} \bigl( A + \sum_{i=1}^5 c_i
u^i \bigr)^2$ for all stable xenon isotopes except $^{131}$Xe~$^\text{a}$, with $u=q^2b^2/2$. The fit function corresponds to the analytical solution given in Refs.~\cite{Donnelly, Lunardini}.\label{tab:fits}}
\begin{center}
\begin{tabular}{l|c|c|c|c|c|c|c}
\hline\hline
Isotope	& $^{128}$Xe & $^{129}$Xe & $^{130}$Xe & $^{131}$Xe 
\footnote{For $^{131}$Xe the fit function is given by $S_S(u) = \frac{2J+1}{4 \pi} \, e^{-u} \bigl[\bigl( A + \sum_{i=1}^5 c_i
u^i \bigr)^2 +\bigl(\sum_{i=1}^5 d_i
u^i \bigr)^2\bigl]$ with the additional fit coefficients: $d_{1}=2.17510$, 
$d_{2}=-1.25401$, $d_{3}=0.214780$, $d_{4}=-0.0111863$, and $d_5=9.21915\cdot10^{-5}$.} & $^{132}$Xe & $^{134}$Xe & $^{136}$Xe \\
\hline
$J^{\Pi}$ & \multicolumn{1}{c|}{$0^+$} & \multicolumn{1}{c|}{$1/2^+$} & \multicolumn{1}{c|}{$0^+$} & \multicolumn{1}{c|}{$3/2^+$} & \multicolumn{1}{c|}{$0^+$} & \multicolumn{1}{c|}{$0^+$} & \multicolumn{1}{c}{$0^+$} \\
\hline
$b$~(fm) & \multicolumn{1}{c|}{2.2827} & \multicolumn{1}{c|}{2.2853} & \multicolumn{1}{c|}{2.2879} & \multicolumn{1}{c|}{2.2905} & \multicolumn{1}{c|}{2.2930} & \multicolumn{1}{c|}{2.2981} & \multicolumn{1}{c}{2.3031} \\
\hline
$c_1$ & $-126.477$ & $-128.119$ & $-129.762$ & $-131.284$ & $-132.841$ & $-135.861$ & $-138.793$\\
$c_2$ & $35.8755$ & $36.5224$ & $37.2824$ & $37.9093$ & $38.4859$ & $39.6999$ & $40.9232$\\
$c_3$ & $-3.71573$ & $-3.8279$ & $-3.94541$ & $-4.05914$ & $-4.08455$ & $-4.2619$ & $-4.43581$\\
$c_4$ & $0.138943$ & $0.152667$ & $0.158662$ & $0.172425$ & $0.153298$ & $0.163642$ & $0.169986$\\
$c_5$ & $-0.00188269$ & $-0.00287012$ & $-0.00288539$ & $-0.00386294$ & $-0.0013897$ & $-0.00164356$ & $-0.00148137$\\
\hline\hline
\end{tabular}
\end{center}
\end{table*}

Because of angular momentum coupling, only $L=0$ multipoles contribute
to the structure factors of the even-mass isotopes. As discussed in
Sec.~\ref{Sec:SI_interaction}, parity and time reversal constrain the
multipoles to even $L$ for elastic scattering, so that for $^{129}$Xe
only $L=0$, and for $^{131}$Xe only $L=0,2$ contribute. For the latter
isotope, we show in Fig.~\ref{fig:Xe131_sf_l} the separate
contributions from $L=0$ and $L=2$ multipoles.  At low momentum
transfers, which is the most important region for experiment, the
$L=0$ multipole is dominant, because coherence is lost for $L>0$
multipoles.  Only near the minima of the $L=0$ multipole at $u \sim
1.7$ and $u \sim 4.4$ is the $L=2$ multipole relevant, but the
structure factor at these $u$ values is suppressed with respect to
$S_S(0)$ by over four and six orders of magnitude, respectively.

Finally, we list in Table~\ref{tab:fits} the coefficients of the fits
performed to reproduce the calculated structure factors for each
isotope.

\begin{figure}[!tb]
\begin{center}
\includegraphics[width=\columnwidth,clip=]{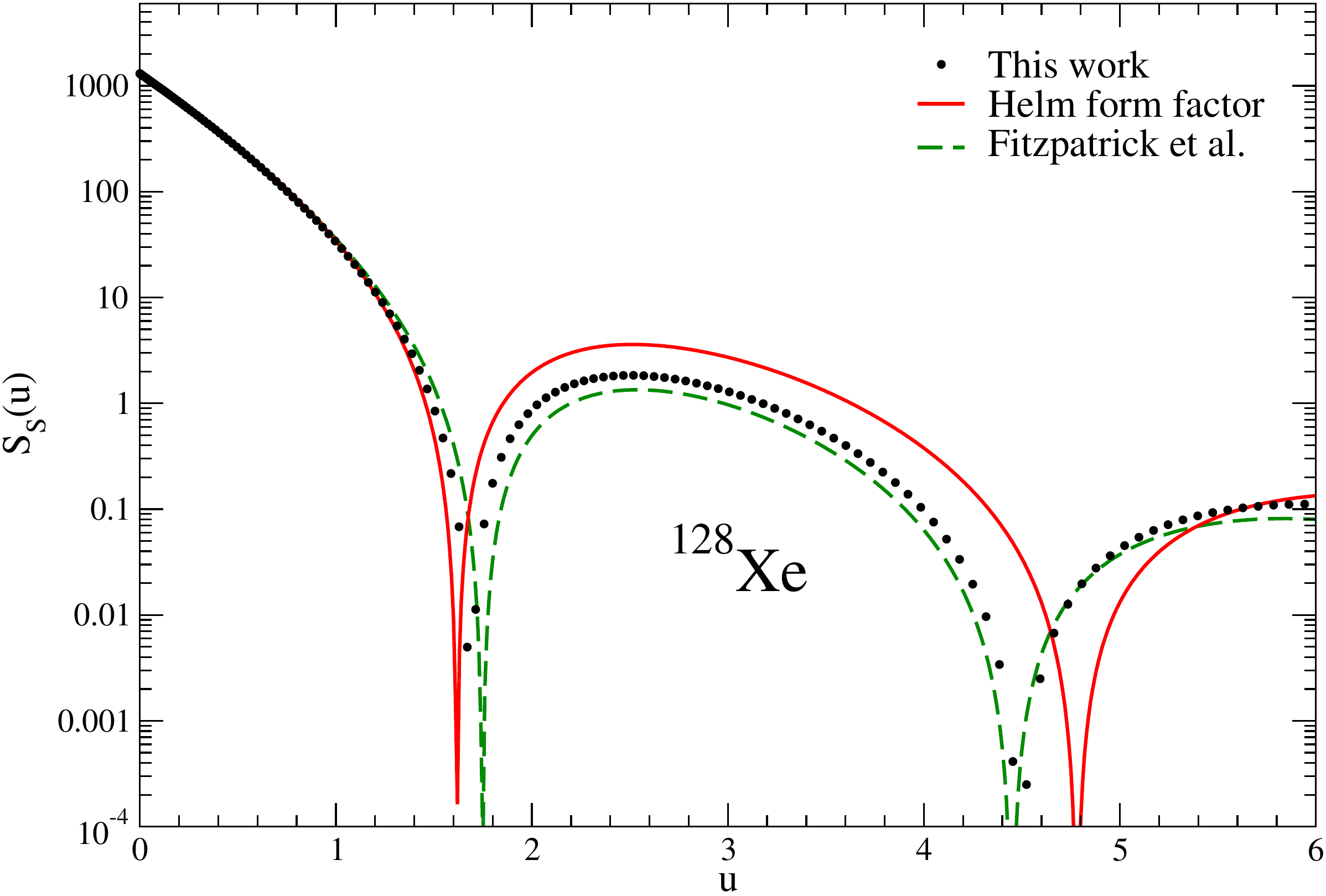}
\end{center}
\caption{(color online). Structure factor $S_S(u)$ for $^{128}$Xe
(this work, black dots) in comparison to the Helm form factor 
(solid red line)~\cite{Lewin} and to the structure factor from
Fitzpatrick {\it et al.} (dashed green line)~\cite{Fitzpatrick}.
\label{fig:Xe128_sf}}
\end{figure}
	
\begin{figure}[!tb]
\begin{center}
\includegraphics[width=\columnwidth,clip=]{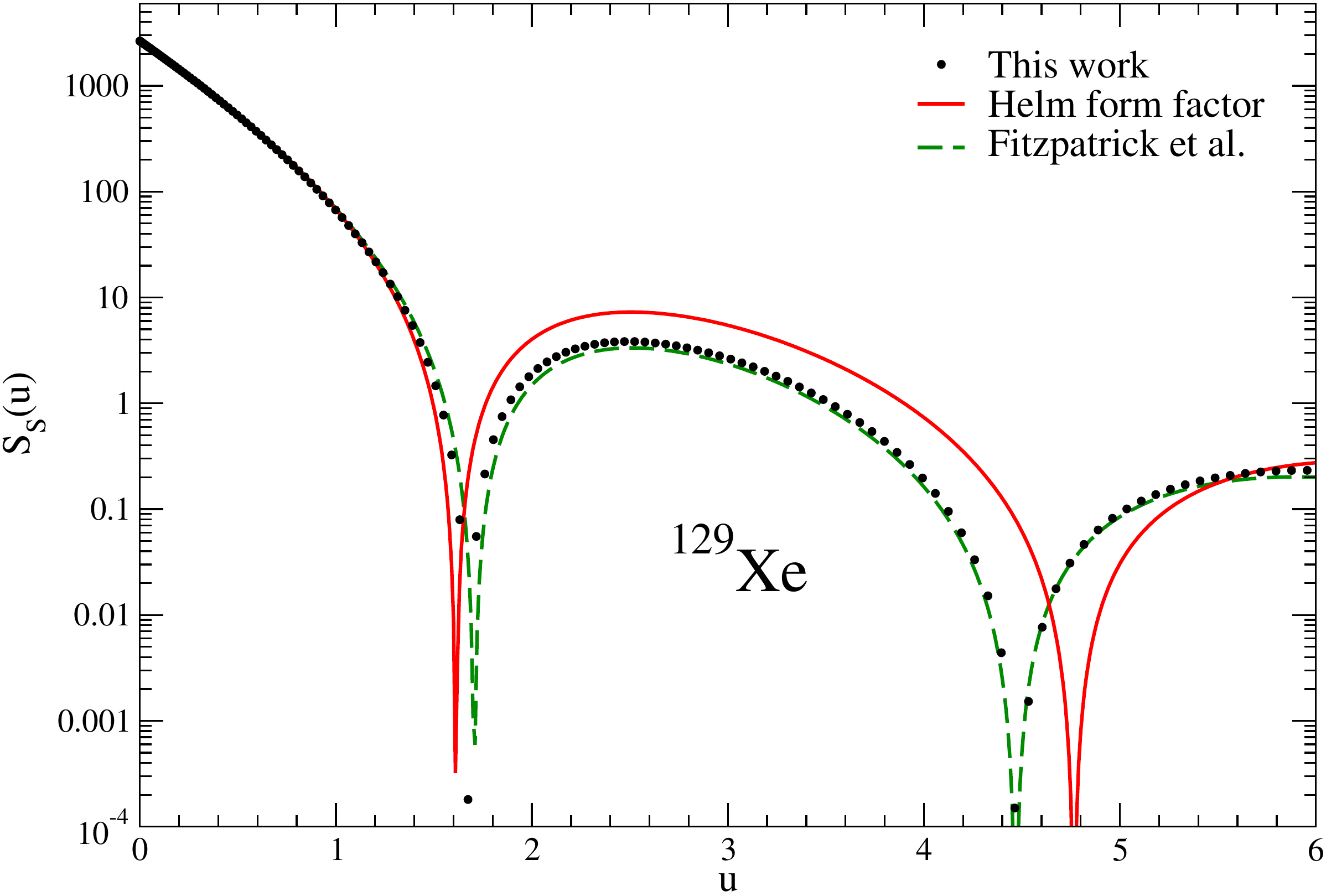}
\end{center}
\caption{(color online). Same as Fig.~\ref{fig:Xe128_sf} but for
$^{129}$Xe.
\label{fig:Xe129_sf}}
\end{figure}

\section{Comparison to Helm form factors and other calculations}
\label{Sec:Comparison Helm}

In experimental SI WIMP scattering analyses the standard structure
factor used to set limits on WIMP-nucleon cross sections is based on
the Helm form factor~\cite{Lewin}. This phenomenological form factor is not
obtained from a detailed nuclear structure calculation, but is based
on the Fourier transform of a nuclear density model, assumed to be
constant with Gaussian surface. The corresponding Helm structure
factor has a simple analytical expression in terms of the nuclear
radius $r_n$ and surface thickness $s$:
\begin{equation}
S_S^\text{Helm}(q) = S_S(0) \,  \biggl( \frac{3j_1(q r_n)}{q r_n} \biggr)^2
e^{-(qs)^2} \,.
\label{eq:Helm}
\end{equation}
The following parameterization is commonly used, $r_n^2 = c^2 +
\frac{7}{3} \pi^2 a^2 - 5 s^2\,$, with $c = (1.23 A^{1/3}-0.60)$~fm,
$a=0.52$~fm, and $s=1$~fm~\cite{Lewin}.

\begin{figure}[t]
\begin{center}
\includegraphics[width=\columnwidth,clip=]{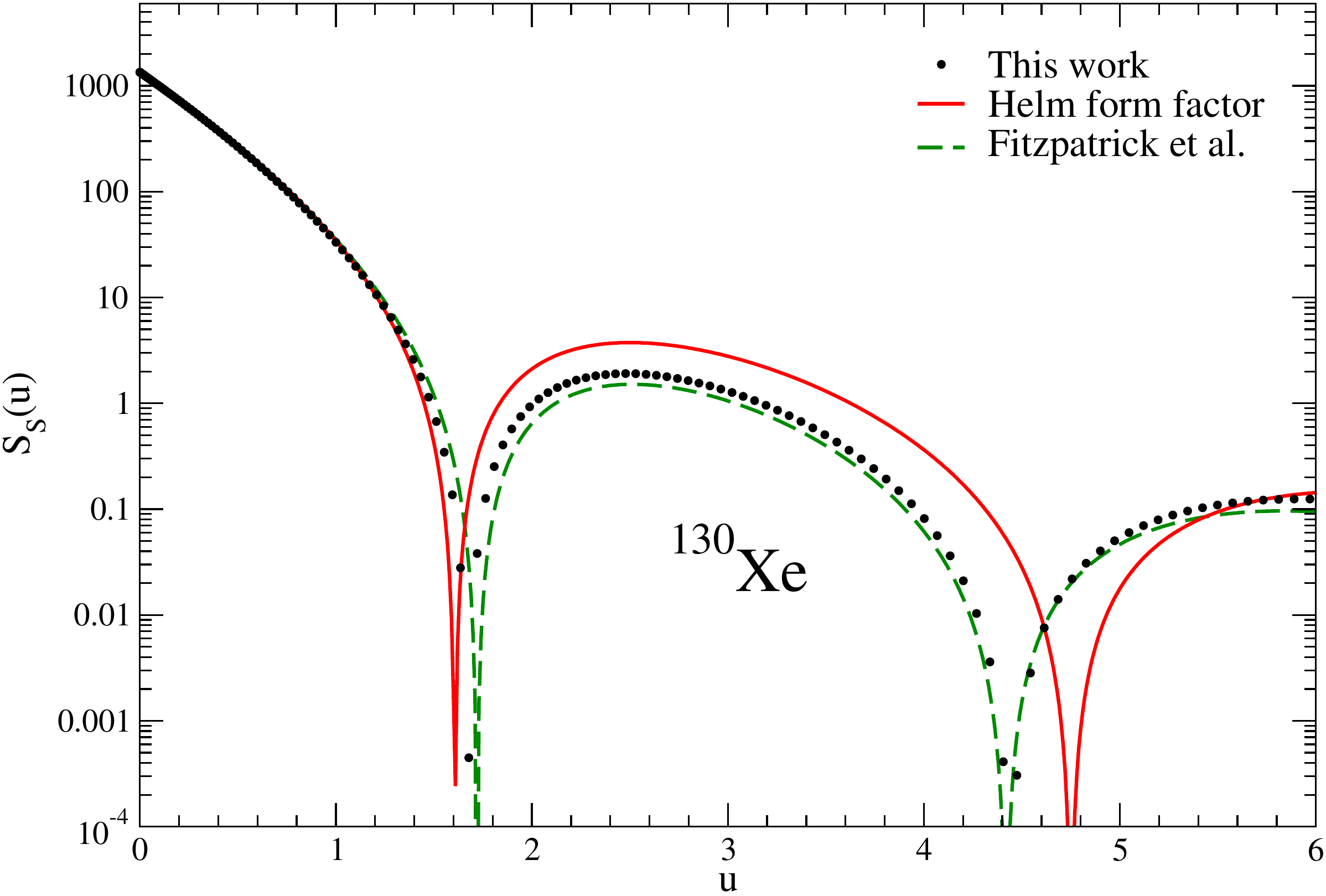}
\end{center}
\caption{(color online). Same as Fig.~\ref{fig:Xe128_sf} but for
$^{130}$Xe.
\label{fig:Xe130_sf}}
\end{figure} 
		
\begin{figure}[t]
\begin{center}
\includegraphics[width=\columnwidth,clip=]{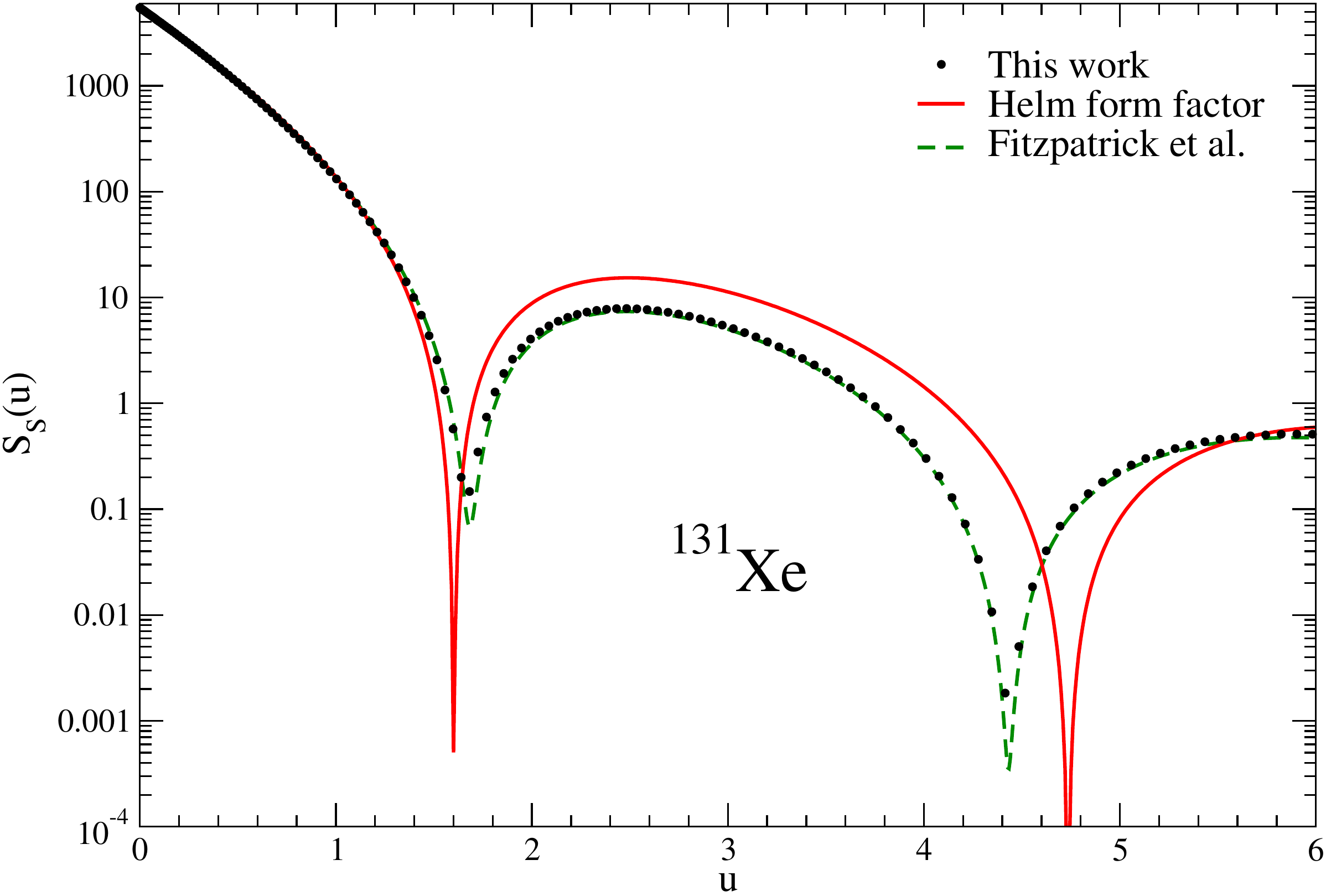}
\end{center}
\caption{(color online). Same as Fig.~\ref{fig:Xe128_sf} but for
$^{131}$Xe.
\label{fig:Xe131_sf}}
\end{figure}

\begin{figure}[t]
\begin{center}
\includegraphics[width=\columnwidth,clip=]{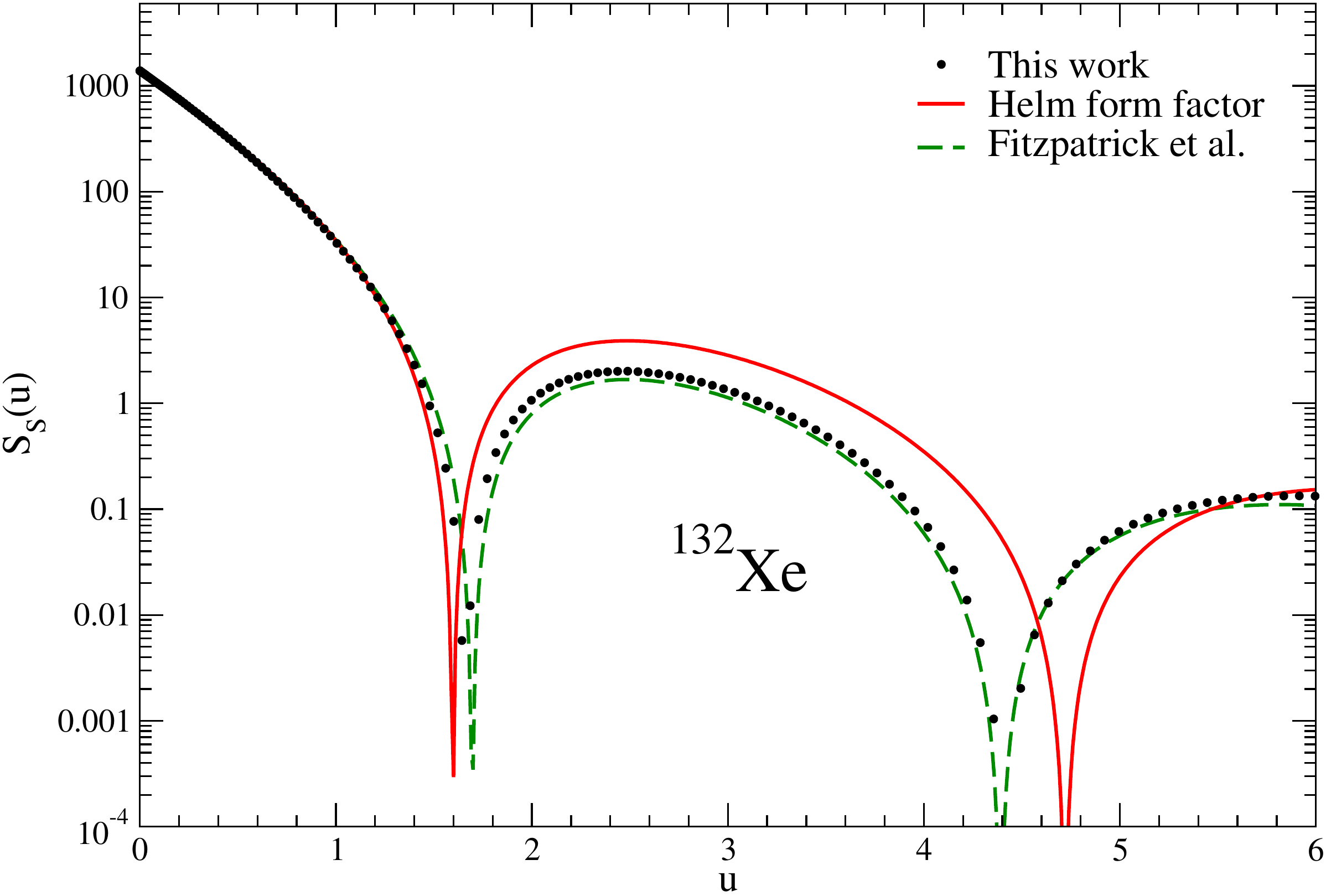}
\end{center}
\caption{(color online). Same as Fig.~\ref{fig:Xe128_sf} but for
$^{132}$Xe.
\label{fig:Xe132_sf}}
\end{figure}

\begin{figure}[t]
\begin{center}
\includegraphics[width=\columnwidth,clip=]{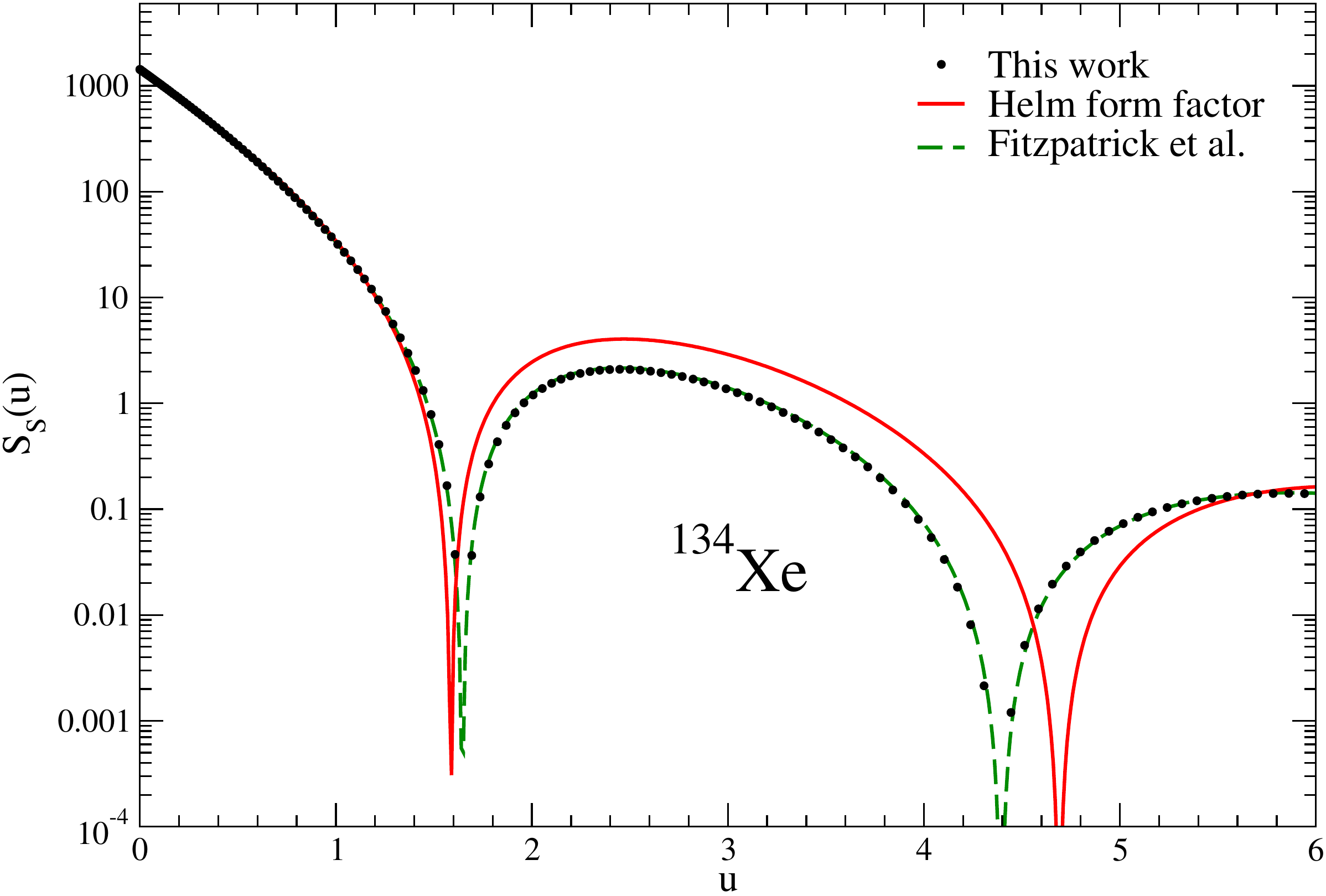}
\end{center}
\caption{(color online). Same as Fig.~\ref{fig:Xe128_sf} but for
$^{134}$Xe.
\label{fig:Xe134_sf}}
\end{figure}

\begin{figure}[!h]
\begin{center}
\includegraphics[width=\columnwidth,clip=]{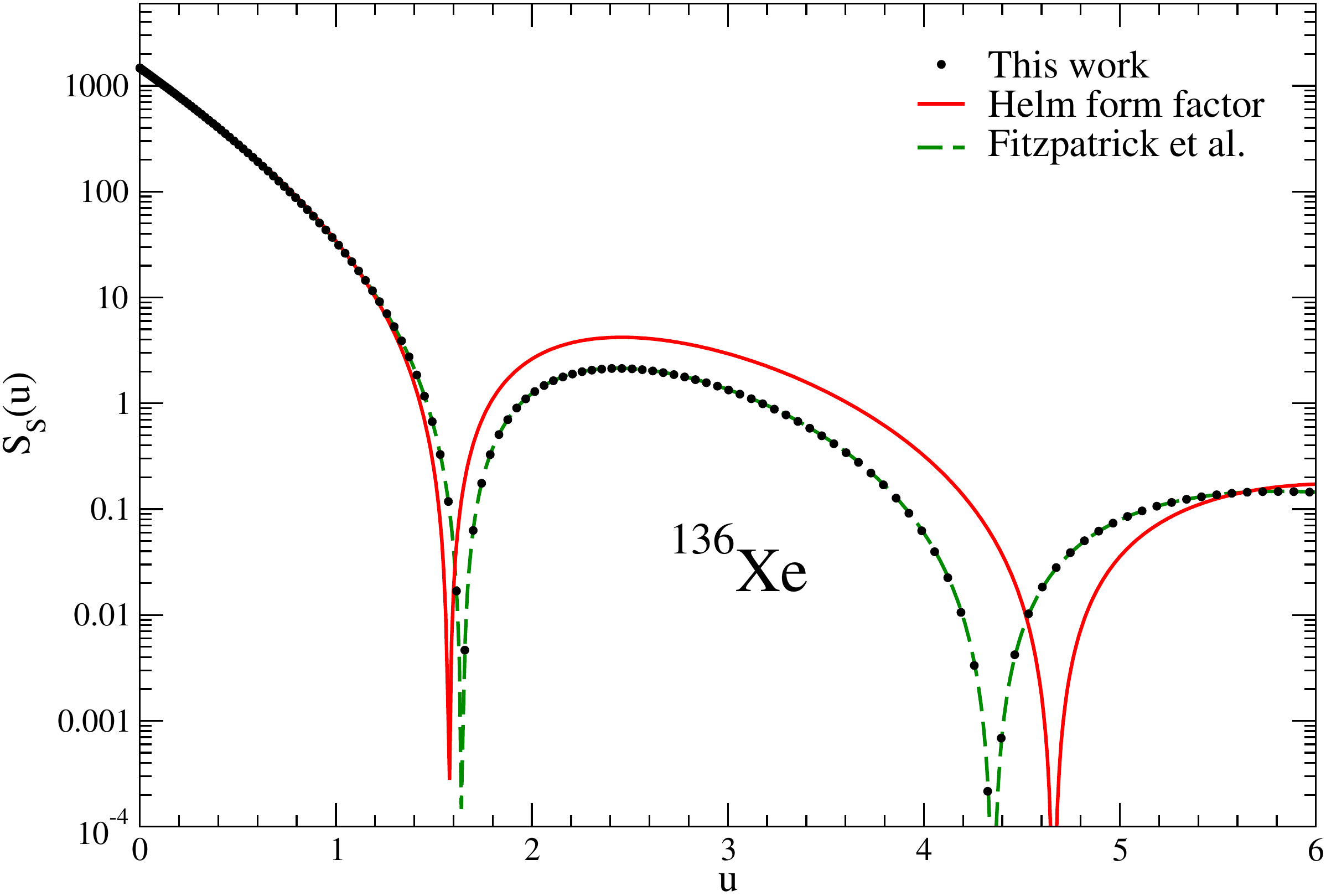}
\end{center}
\caption{(color online). Same as Fig.~\ref{fig:Xe128_sf} but for
$^{136}$Xe.
\label{fig:Xe136_sf}}
\end{figure}

In Figs.~\ref{fig:Xe128_sf}--\ref{fig:Xe136_sf}, we compare the
results for the structure factors presented in Sec.~\ref{sec:structure
  factors} to the phenomenological Helm form
factors given by Eq.~(\ref{eq:Helm}).
At low momentum transfers (and considering
one-body currents only) the agreement is very good for all xenon
isotopes. This validates the present use of Helm form factors in
experimental SI analyses. Similar agreement is expected for other
nuclei considered for WIMP-nucleus scattering.

The first minimum in $S_S(u)$, whose location is set by the nuclear
radius, lies very close in our calculations and the Helm form factors.
At higher momentum transfers small differences start to arise.  The
Helm form factors lie somewhat above our calculations and have the
second minimum at larger momentum transfers. We attribute these minor
differences to the simple assumptions in the Helm form factors.

Figures~\ref{fig:Xe128_sf}--\ref{fig:Xe136_sf} also compare our
results to the structure factors calculated by Fitzpatrick {\it
et al.}~\cite{Fitzpatrick}. These shell-model calculations have been
performed in the same valence space as in our work, but use an older
nuclear interaction and restrict the configurations more severely than
in our case (e.g., only $^{134}$Xe and $^{136}$Xe could be calculated
in the full valence space). Nevertheless, the agreement between the
structure factors of Ref.~\cite{Fitzpatrick} and our present
calculations is very good up to high momentum transfers.  This shows
that at this level SI WIMP scattering is not very sensitive to nuclear
structure details of the isotopes involved. This
conclusion was also reached for even-even nuclei based on
Hartree-Fock calculations~\cite{Co}.

\subsection{Comparison for spin-dependent WIMP scattering}

The interaction of WIMPs with nuclei can be also SD reflecting the
coupling of the spin of the WIMP to nucleons.  The even-mass xenon
isotopes are practically insensitive to SD scattering due to their
$J=0$ ground state, so that only the odd-mass xenon isotopes
$^{129}$Xe and $^{131}$Xe are relevant.  In previous
work~\cite{Menendez,Klos}, we have calculated SD structure factors for
xenon, also including two-body currents in chiral effective field
theory.  To complete the study of WIMP scattering off xenon, we also
compare these calculations to the results obtained by Fitzpatrick
{\it et al.} in Ref.~\cite{Fitzpatrick}. This provides a test of the
calculations and explores the sensitivity of SD WIMP scattering to
nuclear structure.

The SD structure factor is naturally decomposed in terms of the isospin couplings $(a_0 + a_1 \tau_3)/2$. However,
experimental results are commonly presented in terms of
``neutron-only'' ($a_0=-a_1=1$) and ``proton-only'' ($a_0=a_1=1$)
structure factors $S_n(u)$ and $S_p(u)$, because these coupling
combinations are more sensitive to neutrons and protons, respectively.
For vanishing momentum transfer, $q=0$ ($u=0$), and considering only
one-body currents, the SD ``neutron-only'' and ``proton-only''
structure factors are proportional to the square of the expectation
values of the neutron and proton spins~\cite{Engel}. These are given
for both calculations in Table~\ref{tab:spin exp value}. Because xenon
has an even proton number, $\langle \mathbf{S}_n \rangle \gg \langle
\mathbf{S}_p \rangle$, the ``neutron-only'' structure factor
dominates over the ``proton-only'' one.

\begin{table}[b]
\caption{Proton/neutron spin expectation values
$\langle \mathbf{S}_{p/n} \rangle$ for $^{129}$Xe and $^{131}$Xe.
Results are shown for the calculations of 
Klos {\it et al.}~\cite{Klos}, which use the same valence space
and nuclear interactions as in this work, and of Fitzpatrick
{\it et al.}~\cite{Fitzpatrick}.\label{tab:spin exp value}}
\begin{center}
\begin{tabular}{l|c|c|c|c}
\hline\hline
& \multicolumn{2}{c|}{$^{129}$Xe} & \multicolumn{2}{c}{$^{131}$Xe} \\
\hline
& $\langle \mathbf{S}_p \rangle$ & $\langle \mathbf{S}_n \rangle$ 
& $\langle \mathbf{S}_p \rangle$ & $\langle \mathbf{S}_n \rangle$ \\
\hline
Klos {\it et al.}~\cite{Klos}\, & \,$0.010$\,& \,$0.329$\,& \,$-0.009$\,& \,$-0.272$ \\
Fitzpatrick {\it et al.}~\cite{Fitzpatrick}\, & \,$0.007$\, & \,$0.248$\, & \,$-0.005$ \,  &\,$-0.199$ \\
\hline\hline
\end{tabular}
\end{center}
\end{table}

\begin{figure}[t]
\begin{center}
\includegraphics[width=\columnwidth,clip=]{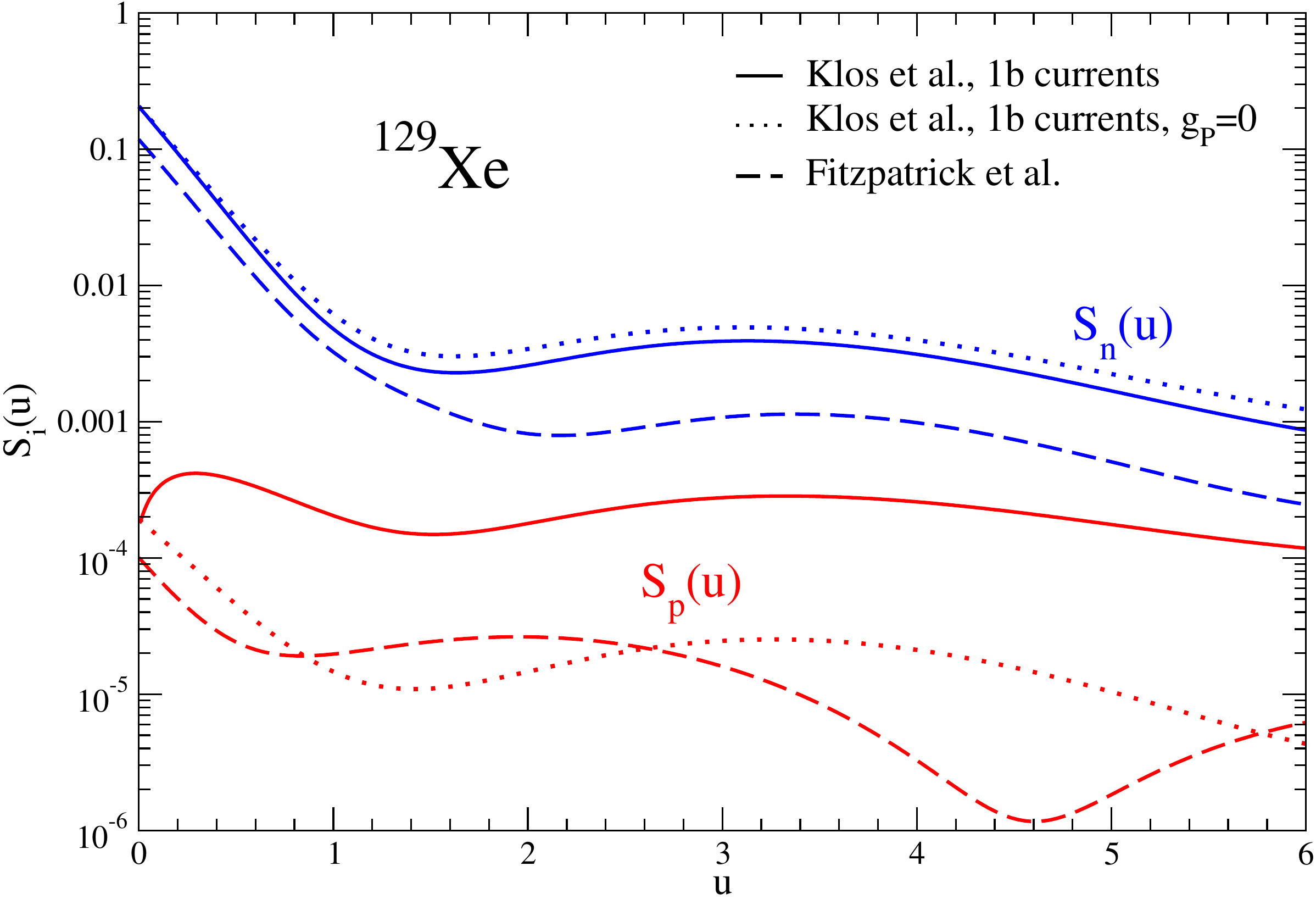}
\end{center}
\caption{(color online). Comparison of ``neutron-only'' $S_n(u)$ (blue)
and  ``proton-only'' $S_p(u)$ (red lines) spin-dependent structure factors for
$^{129}$Xe: results are shown from Klos {\it et al.}~\cite{Klos} at
the one-body (1b) current level with/without pseudoscalar ($g_P$)
contributions (solid/dotted lines) and from Fitzpatrick 
{\it et al.}~\cite{Fitzpatrick} (dashed lines).
\label{fig:Xe129_sd}}
\end{figure}

\begin{figure}[t]
\begin{center}
\includegraphics[width=\columnwidth,clip=]{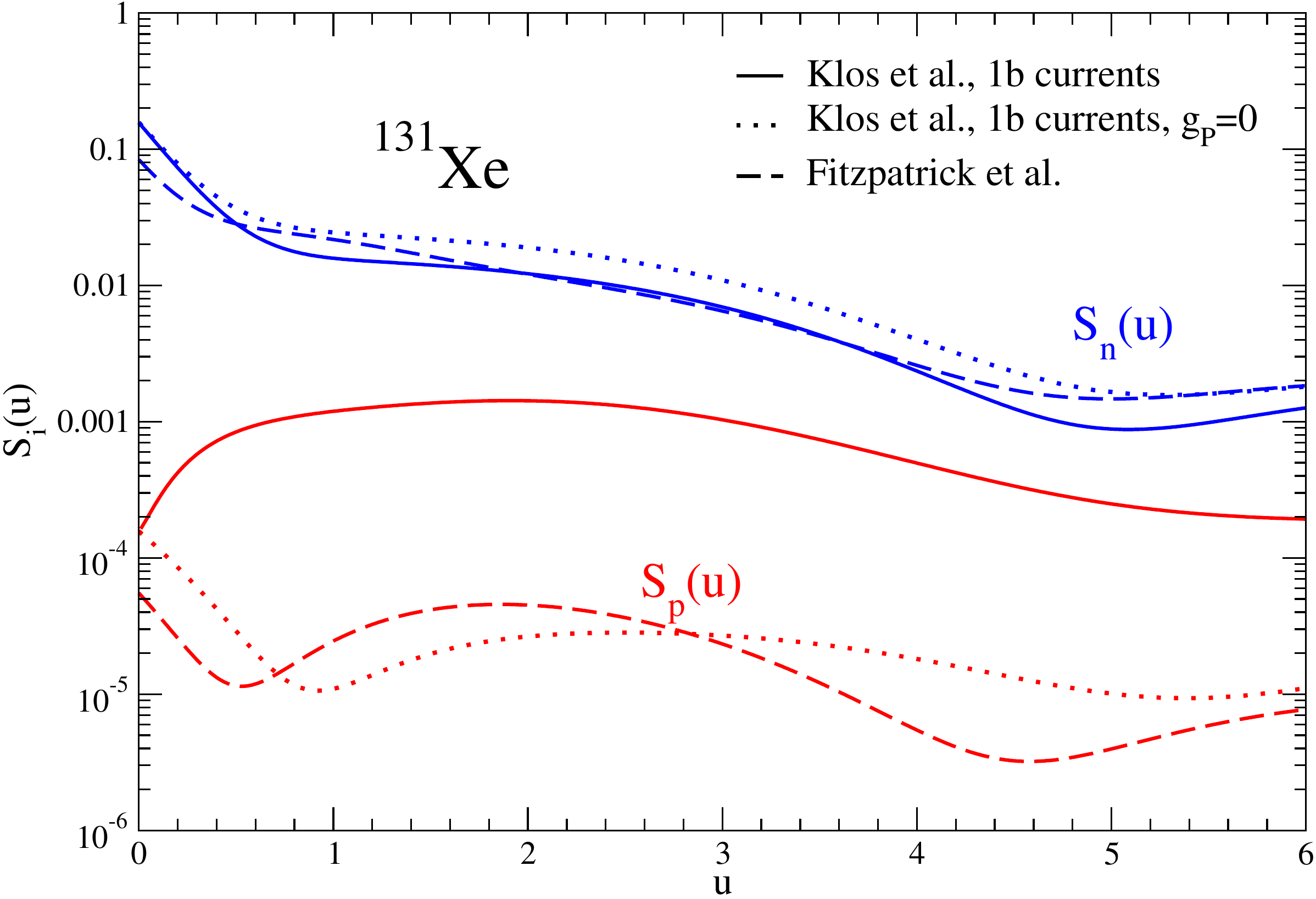}
\end{center}
\caption{(color online). Same as Fig.~\ref{fig:Xe129_sd} but for
$^{131}$Xe.
\label{fig:Xe131_sd}}
\end{figure}

This hierarchy of ``neutron-only'' versus ``proton-only'' structure
factors manifests itself in Figs.~\ref{fig:Xe129_sd}
and~\ref{fig:Xe131_sd}, where we show the calculated SD structure
factors for $^{129}$Xe and $^{131}$Xe. Note that the absolute scale of
the SD structure factors is $\sim 10^{-4}$ smaller than for SI
scattering, because in the SD case, due to pairing, the contributions
from different nucleons do not add coherently.
	
In Refs.~\cite{Menendez,Klos}, we included one- and two-body currents
in the WIMP-nucleon interaction Lagrangian.  However, for a direct
comparison, Figs.~\ref{fig:Xe129_sd} and~\ref{fig:Xe131_sd} restrict
the results to the one-body level, even though two-body currents are
important because they reduce the ``neutron-only'' structure factors
by about $20\%$ for xenon, and significantly enhance the
``proton-only'' structure factors at small momentum
transfers~\cite{Menendez,Klos}.  In addition to the structure factors
calculated with the full one-body currents, Fig.~\ref{fig:Xe129_sd}
and Fig.~\ref{fig:Xe131_sd} also show results without the pseudoscalar
contributions ($g_P=0$).  This choice can be directly compared to the operator used by Fitzpatrick {\it et al.}~\cite{Fitzpatrick}, because the pseudoscalar contributions are considered as an independent response~\cite{Fitzpatrick}.

A comparison of the different calculations in Figs.~\ref{fig:Xe129_sd}
and~\ref{fig:Xe131_sd} shows larger differences than for SI
scattering.  This is because the SD case is more sensitive to nuclear
structure details.  At $u=0$ this difference can be
traced to the larger spin expectation value obtained in Klos 
{\it et al.}~\cite{Klos} (see Table~\ref{tab:spin exp value}), 
which is a modest 1.3 times larger than in Fitzpatrick 
{\it et al.}~\cite{Fitzpatrick} for both isotopes. The difference is 
due to the more recent nuclear interaction used with less
truncations, and also explains why this ``neutron-only'' response is
larger for any $u$ value.  Figures~\ref{fig:Xe129_sd}
and~\ref{fig:Xe131_sd} also show that, for finite momentum transfer,
the pseudoscalar contributions enhance the ``proton-only'' structure
factor by about one order of magnitude.  This is because when this
isovector term is included, neutrons, which carry most of the spin in
xenon, can contribute to the ``proton-only'' response.
Because the relative strength of axial and pseudoscalar contributions in the Fitzpatrick {\it et al.}\ calculation~\cite{Fitzpatrick} are taken to be a ratio of independent couplings, we compare the two calculations in the limit where the pseudoscalar coupling is turned off. In this limit, the results are comparable at higher momentum transfers and the differences are similar as for the neutron-only case.

We emphasize that the two shell-model calculations agree in the sign
and magnitude of the matrix element ratios $\langle {\bf S}_p \rangle /
\langle {\bf S}_n \rangle \sim 0.03$ for both isotopes, so that the proton
amplitude is about 3\% of the total. Moreover, the agreement between
structure factors in the physically relevant region $u \lesssim 1$ is better
than at high-momentum transfers, where other corrections not included
in these calculations will be relevant. This shows that the
uncertainties in the structure factors are modest, so that they should
not limit the extraction of dark matter information from direct
detection experiments (see also the conclusions of Ref.~\cite{Cerdeno}).

\section{Spin-independent vs. spin-dependent inelastic scattering}
\label{Sec:Spin-independent}

The xenon isotopes $^{129}$Xe and $^{131}$Xe have $J_f=3/2^+$ and
$J_f=1/2^+$ low-lying excited states at $39.6$~keV and and $80.2$~keV, respectively,
that could be excited in inelastic WIMP scattering. In
Ref.~\cite{inelastic} we showed that for these isotopes, at the
momentum transfers kinematically allowed for inelastic scattering
(corresponding to $u\sim1$), the SD elastic and inelastic structure
factors are comparable, and the inelastic maxima are suppressed by
only a factor 10 compared to the elastic case.  This opens the door to
the detection of SD inelastic WIMP scattering off xenon. Note that
elastic scattering is always dominant because of its maximum at $q=0$
and more favorable kinematics.

\begin{figure}[t]
\begin{center}
\includegraphics[width=\columnwidth,clip=]{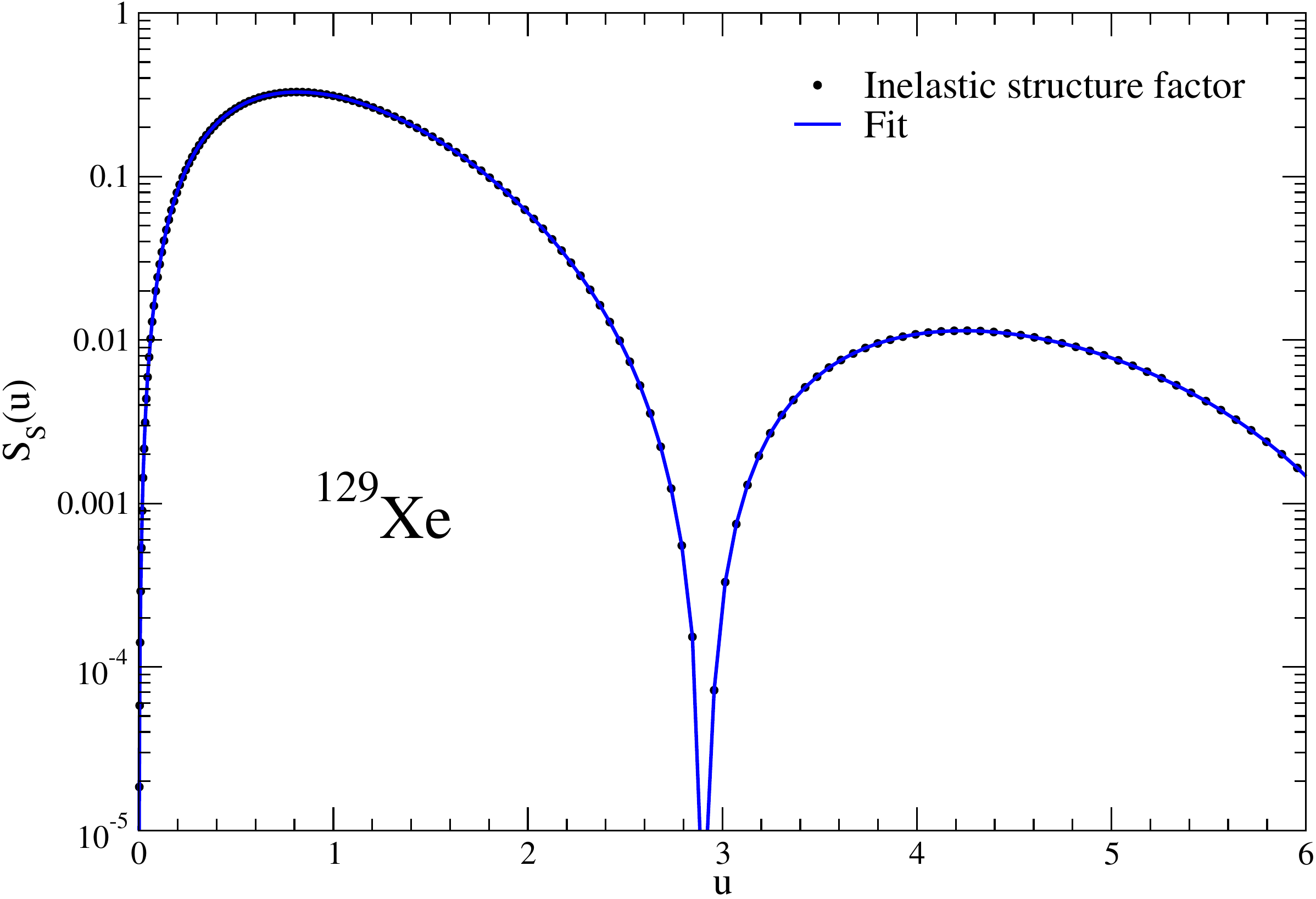}
\end{center}
\caption{(color online). Spin-independent inelastic structure factor
for $^{129}$Xe (black dots), from the $J_i=1/2^+$ ground state to the
$J_f=3/2^+$ excited state at $39.6$~keV, with a fit (solid blue line)
given in Table~\ref{tab:fits_inelastic}.\label{fig:Xe129_sf_inelastic}}
\end{figure}

\begin{figure}[t]
\begin{center}
\includegraphics[width=\columnwidth,clip=]{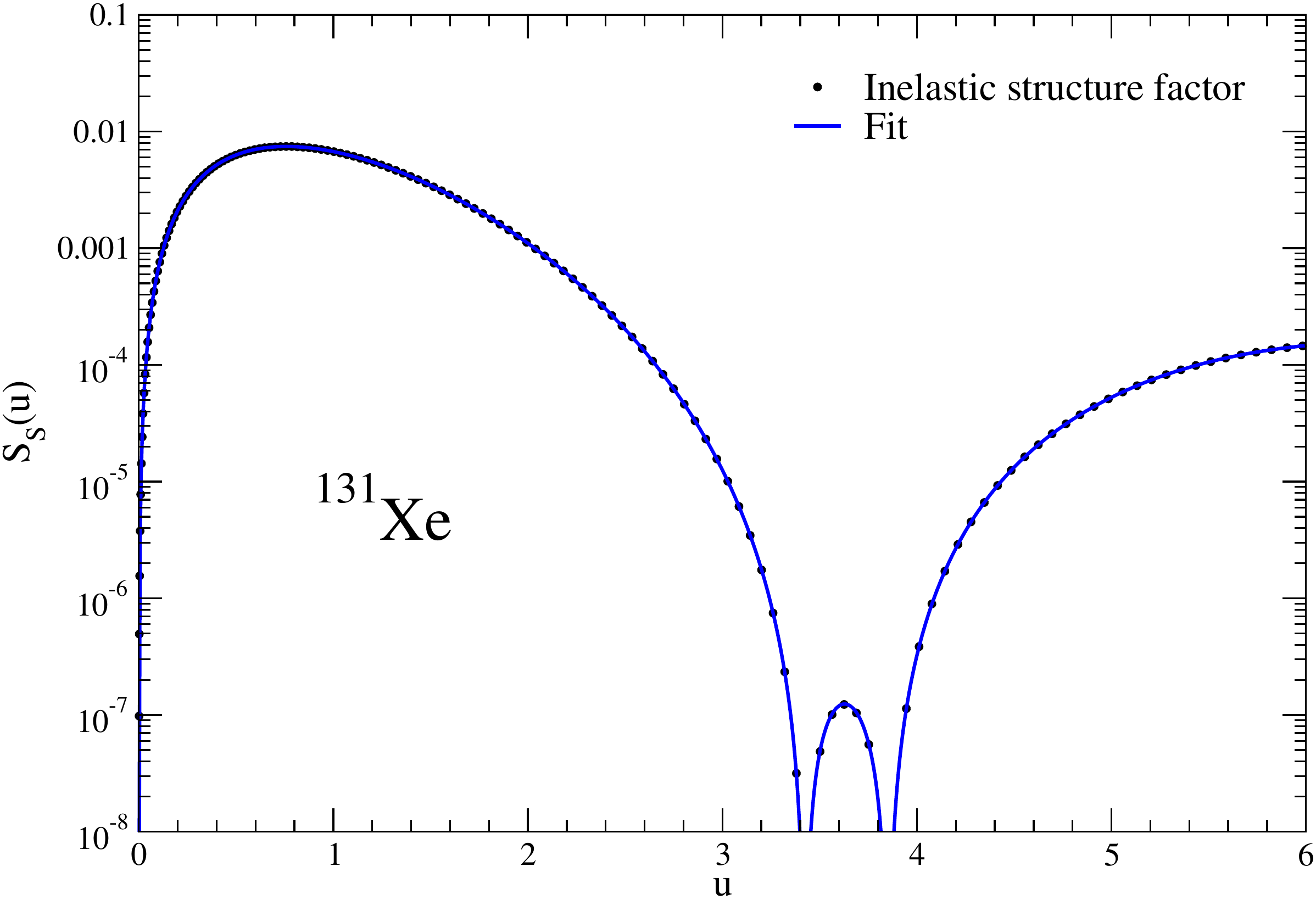}
\end{center}
\caption{(color online). Spin-independent inelastic structure factor
for $^{131}$Xe (black dots), from the $J_i=3/2^+$ ground state to the
$J_f=1/2^+$ excited state at $80.2$~keV, with a fit (solid blue line)
given in Table~\ref{tab:fits_inelastic}.\label{fig:Xe131_sf_inelastic}}
\end{figure}

Figures~\ref{fig:Xe129_sf_inelastic} and~\ref{fig:Xe131_sf_inelastic}
show the calculated structure factors for SI inelastic scattering to
the lowest excited state in $^{129}$Xe and $^{131}$Xe.  Angular
momentum and parity considerations limit the multipole contributions
to $L=2$ in both cases.  For SI inelastic scattering, the
contributions from different nucleons do not add coherently, and the
structure factors are suppressed by several orders of magnitude with
respect to the elastic case.  At the kinematically allowed region for
inelastic scattering around $u \sim 1$, the suppression is about a
factor $2 \cdot 10^{-3}$ for $^{129}$Xe and $10^{-4}$ for
$^{131}$Xe. When comparing the global maxima for elastic and inelastic scattering,
the suppression is even stronger, by factors of about
$10^{-4}$ and $5 \cdot 10^{-5}$, respectively, in stark contrast to SD
scattering.

\begin{table}[h!]
\caption{Fit coefficients for the inelastic structure factors $S_S(u)$
corresponding to $S_S(u) = \frac{2J_i+1}{4 \pi} \, e^{-u} (\sum_{i=1}^5 
d_i u^i)^2$ for $^{129}$Xe and $^{131}$Xe, with $u=q^2b^2/2$ and $J_i=1/2^+$
and $J_i=3/2^+$, respectively. The fit function corresponds to the analytical solution given in Refs.~\cite{Donnelly, Lunardini}. The harmonic-oscillator lengths $b$ are
as in Table~\ref{tab:fits}.\label{tab:fits_inelastic}}
\begin{center}
\begin{tabular}{l|c|c}
\hline\hline
Isotope	& \multicolumn{1}{c|}{$^{129}$Xe} & \multicolumn{1}{c}{$^{131}$Xe} \\
\hline
$d_1$ & $4.46850$ & $0.515046$ \\
$d_2$ & $-2.54918$ & $-0.341605$ \\
$d_3$ & $0.406162$ & $0.0707621$ \\
$d_4$ & $-0.0206094$ & $-0.00436258$ \\
$d_5$ & \,$0.000258314$\, & \,$9.81102\cdot10^{-7}$\, \\
\hline\hline
\end{tabular}
\end{center}
\end{table}

This explicitly confirms that inelastic scattering can discriminate
between SI and SD interactions, as detection of the inelastic channel
would point to a SD WIMP-nucleus coupling~\cite{inelastic}. Finally,
the fit coefficients for the SI inelastic structure factors for
$^{129}$Xe and $^{131}$Xe are listed in Table~\ref{tab:fits_inelastic}.

\section{Summary}
\label{Sec:Summary}

We have studied SI WIMP scattering off xenon using the leading
one-body scalar currents. Our nuclear structure calculations are based
on state-of-the-art shell-model calculations in the largest valence
spaces with interactions that have been tested against spectroscopy
and decay studies. In particular, the spectra of all relevant xenon
isotopes are very well reproduced.

Based on these nuclear interactions, we have calculated the structure
factors for the xenon isotopes. These present the consistent
calculations to the SD results in Refs.~\cite{Klos}, providing fits
for all structure factors.  For the momentum transfers relevant to
direct detection experiments, $u \lesssim 1$, the calculated structure
factors are in very good agreement with the phenomenological Helm form
factors used to give experimental limits for dark matter
detection. This shows that the presently extracted limits from
SI~\cite{XenonSI,LUX} and SD~\cite{XenonSD} interactions off xenon are
consistent in the underlying nuclear structure used for the analysis.

In addition we have compared our results for the structure factors to
the shell-model calculations of Fitzpatrick {\it et al.}~\cite{Fitzpatrick},
which have been performed with more truncations and older nuclear
interactions. However, because SI scattering is sensitive to the
nucleon density distribution, both calculations agree well. In
particular, for $u \lesssim 1$, the agreement is excellent. This shows
that the spin-independent structure factor is not very sensitive to
details in nuclear interactions. However, we emphasize that additional
contributions are expected from two-body
currents~\cite{Prezeau,Cirigliano}.

In contrast, for SD interactions, even at the one-body level, there
are larger differences between the results of Klos 
{\it et al.}~\cite{Klos}, which use the same nuclear interactions as
in this work, and those of Fitzpatrick {\it et  al.}~\cite{Fitzpatrick}.
These differences are mostly due to the different spin expectation values at $u=0$.
However, these differences are modest and should not limit
the extraction of dark matter information from direct detection
experiments. Efforts to further reduce the uncertainties based on
nuclear structure input are underway.

Finally, we have calculated the structure factors for SI inelastic
scattering for the odd-mass xenon isotopes $^{129}$Xe and
$^{131}$Xe. These have low-lying excited states that can be accessed
by WIMP scattering~\cite{inelastic}. As expected, the inelastic
response is suppressed by $\sim 10^{-4}$ compared to coherent elastic
scattering. Therefore, the detection of inelastic scattering is able
to discriminate clearly between SI and SD scattering, because the SD
inelastic structure factor, while suppressed relative to elastic
scattering at $u=0$, becomes comparable for $u \sim 1$, where the
inelastic response is suppressed only by a factor 10 with respect to
the elastic maximum. This demonstrates how using nuclear properties
will be important for decoding the information from dark matter
signals.

\section*{Acknowledgments}

We thank M.\ Hoferichter for discussions. This work was supported by
ARCHES, the DFG through Grant No.~SFB 634, the ERC Grant No.~307986
STRONGINT, the Helmholtz Alliance Program of the Helmholtz
Association, contract HA216/EMMI ``Extremes of Density and
Temperature: Cosmic Matter in the Laboratory'', and by the US
Department of Energy under Contracts DE-SC00046548 and
DE-AC02-98CH10886. W.\ H.\ thanks the Alexander von Humboldt
Foundation for support.

\end{document}